\documentclass[pdflatex,sn-basic,iicol]{sn-jnl}% Basic Springer Nature Reference Style/Chemistry Reference Style

\jyear{2023}%
\usepackage{amsmath}
\usepackage[version=4]{mhchem}
\usepackage{bm}
\usepackage{cancel}
\usepackage{times}

\theoremstyle{thmstyleone}%
\theoremstyle{thmstyletwo}%

\theoremstyle{thmstylethree}%

\begin{document}

\title[Article Title]{\textbf{Force moment partitioning and scaling analysis of vortices shed by a 2D pitching wing in quiescent fluid}}

\author*[1]{\fnm{Yuanhang} \sur{Zhu}}\email{yuanhang\_zhu@brown.edu}\presentaddressone{Department of Mechanical and Aerospace Engineering, University of Virginia, Charlottesville, VA 22904, USA}

\author[1]{\fnm{Howon} \sur{Lee}}\presentaddresstwo{School of Aerospace Engineering, Georgia Institute of Technology, Atlanta, GA 30332, USA}

\author[2]{\fnm{Sushrut} \sur{Kumar}}

\author[3]{\fnm{Karthik} \sur{Menon}}

\author[2]{\fnm{Rajat} \sur{Mittal}}

\author[1]{\fnm{Kenneth} \sur{Breuer}}

\affil[1]{\orgdiv{Center for Fluid Mechanics, School of Engineering}, \orgname{Brown University}, \orgaddress{\city{Providence}, \state{RI}, \postcode{02912}, \country{USA}}}

\affil[2]{\orgdiv{Department of Mechanical Engineering}, \orgname{Johns Hopkins University}, \orgaddress{\city{Baltimore}, \state{MD}, \postcode{21218}, \country{USA}}}

\affil[3]{\orgdiv{Department of Pediatrics (Cardiology) and Institute for Computational and Mathematical Engineering}, \orgname{Stanford University}, \orgaddress{\city{Stanford}, \state{CA}, \postcode{94305}, \country{USA}}}

\abstract{We experimentally study the dynamics and strength of vortices shed from a NACA 0012 wing undergoing sinusoidal pitching in quiescent water. We characterize the temporal evolution of the vortex trajectory and circulation over a range of pitching frequencies, amplitudes and pivot locations. By employing a physics-based force and moment partitioning method (FMPM), we estimate the vortex-induced aerodynamic moment from the velocity fields measured using particle image velocimetry (PIV). The vortex circulation, formation time and vorticity-induced moment are shown to follow scaling laws based on the feeding shear-layer velocity. The vortex dynamics, together with the spatial distribution of the vorticity-induced moment, provide quantitative explanations for the nonlinear behaviors observed in the fluid damping (Zhu \emph{et al.}, \emph{J. Fluid Mech.}, vol. 923, 2021, R2). The FMPM-estimated moment and damping are shown to match well in trend with direct force measurements, despite a discrepancy in magnitude. Our results demonstrate the powerful capability of the FMPM in dissecting experimental flow field data and providing valuable insights into the underlying flow physics.}

\maketitle

\section{Introduction}\label{sec.intro}

The unsteady flow and vortex dynamics associated with flapping wings/foils have been extensively studied for understanding the complex lift/thrust generation mechanism of animal flight and swimming \citep{ellington1996leading,lentink2009rotational,anderson1998oscillating,triantafyllou2000hydrodynamics}, as well as developing bio-inspired flapping-wing micro air vehicles \citep[MAVs,][]{ho2003unsteady,shyy2010recent,jafferis2019untethered}, oscillating-foil autonomous underwater vehicles \citep[AUVs,][]{barrett1996propulsive,zhu2019tuna,zhong2021tunable} and energy-harvesting devices \citep{xiao2014review,young2014review}. Flapping wings/foils were commonly studied in the presence of an ambient flow to match real flying/swimming settings. Numerous studies have focused on identifying scaling laws that characterize the formation and development of the shed vortices, along with the corresponding aerodynamic loads \citep{buchholz2008wake,buchholz2011scaling,baik2012unsteady,onoue2016vortex}. In the absence of a freestream flow, the problem is associated with the hovering flight of insects \citep{wang2005dissecting,bergou2007passive,kang2014analytical} and the starting motion of fish \citep{heathcote2004flexible,epps2007impulse,devoria2012vortex,devoria2013force,shinde2013jet,david2018kinematic}. In these studies, the pitching panel was usually hinged at the leading-edge to mimic the corresponding biological appendage, and more emphasis was placed on the vorticity-induced lift and thrust. On the other hand, the vorticity-induced aerodynamic moment has attracted less attention, despite its close connections with maneuvering and its importance in regulating aerodynamic damping.

In aeroelastic systems, the formation and shedding of vortices from elastically mounted bluff bodies is the main source of aerodynamic damping \citep{williamson2004vortex,morse2009prediction,menon2019flow,zhu2020nonlinear}. \citet{menon2019flow} and \citet{zhu2020nonlinear} have analyzed the energy transfer between an elastically mounted pitching wing and the unsteady ambient flow, and found that the total damping in an aeroelastic system has to equal to zero for flow-induced oscillations to sustain. This means that the positive structural damping has to balance the negative aerodynamic damping. With a free-stream flow, the negative aerodynamic damping mainly comes from the dynamic stall vortex \citep{mccroskey1982unsteady,corke2015dynamic}. Without a free-stream flow, however, the aerodynamic damping becomes purely positive, as the pitch-induced vortices act as a source of drag \citep{morison1950force}. \citet{zhu2021nonlinear} have adopted a dynamical system approach to study the nonlinear fluid damping associated with vortices shed from a cyber-physically mounted pitching wing in the absence of a free-stream flow. They extracted the fluid damping coefficient using ``ring-down'' experiments and were able to identify a universal non-dimensional fluid damping coefficient that depends on the pitching frequency, amplitude, pivot location and sweep angle. They also explained the nonlinear behavior of the fluid damping using the corresponding vortex dynamics. However, their analysis of the complex vortex dynamics was purely qualitative, and no quantitative evaluations of the vortex trajectory and strength were performed. Another limitation of \citet{zhu2021nonlinear}'s work was that the dynamical system framework was only capable of resolving cycle-averaged vortex-induced damping. No direct measurements of the instantaneous force and moment associated with the shed vortices were available, although correlating this information with the corresponding vortex dynamics could provide useful insights toward a more complete understanding of the underlying flow physics. 

The recent development of the Force and Moment Partitioning Method \citep[FMPM,][]{quartapelle1983force,zhang2015centripetal,moriche2017aerodynamic,menon2021initiation,menon2021quantitative,menon2021significance} \citep[a variant is also known as the vortex force/moment map method,][]{li2018vortex,li2020vortex} provides us with a framework to determine the instantaneous flow-induced forces and moments from the corresponding velocity fields. In the FMPM, the Navier-Stokes equation is projected onto the gradient of an auxiliary potential which satisfies the Laplace equation and certain boundary conditions. The individual contributions of the added-mass, vorticity-induced, and viscous terms to the fluid force/moment exerted on the immersed body can be separated analytically, enabling independent dissection of each term. Moreover, the FMPM is able to visualize the spatial distribution of the flow-induced force/moment, which is valuable for associating the vortex dynamics with the resultant aerodynamic loads. \citet{zhang2015centripetal} have applied FMPM to a numerical simulation of the flight of a hawkmoth and a fruit fly and found that, in addition to the leading-edge vortex, which contributes to the majority of the lift, the centripetal acceleration reaction, which is analogous to the added-mass effect, also plays an important role in the overall lift generation. \citet{menon2021initiation} have revisited the classical problem of vortex-induced vibrations associated with flow over a cylinder and found using FMPM that the self-sustained oscillations are driven by the vorticity in the shear layer instead of the shed vortices in the wake. Moreover, using FMPM, \citet{menon2021significance} discovered that, in addition to the rotation-dominated region (i.e. regions defined as vortices), the strain-dominated region surrounding the vortices also has a significant effect on the aerodynamic load generation of pitching airfoils. \citet{seo2022improved} have used FMPM in conjunction with direct numerical simulations to show that most of the thrust generated by a carangiform swimmer is due to the leading-edge vortex on the caudal fin.

The above examples have demonstrated the powerful capability of FMPM to provide new physical insights for vortex-dominated flows. However, in these examples, the FMPM was only tested and validated on results obtained from numerical simulations, where data was usually considered to be clean and ideal. Applying FMPM to experimental data will open up numerous new possibilities for flow measurement techniques such as particle image velocimetry (PIV). However, there are many foreseeable challenges. First, in many experiments, the measured velocity fields are usually ensemble- or phase-averaged so as to reduce measurement noise and incoherent flow structures, but it is not clear how ensemble- or phase-averaging will affect the accuracy of the FMPM due to the nonlinear operations involved. Second, it is well-acknowledged that the accuracy of PIV measurements suffers close to the boundary of immersed bodies, which brings challenges to PIV-based load estimation methods \citep{rival2017load}. Third, for nominally two-dimensional flows measured using planar PIV, the existence of out-of-plane flows might introduce errors into the FMPM, depending on the strength of the three-dimensionality of the flow.

The present work extends the study of \citet{zhu2021nonlinear} on cycle-averaged vortex-induced damping and examines the \textit{instantaneous} vortex trajectory, strength, structure, and vorticity-induced moment of sinusoidally pitching wings in a quiescent fluid. We also aim to use the present problem as an example to address the aforementioned potential challenges of applying FMPM to experimental data. Since the current experiment focuses on a pitching wing, we confine our analysis to \emph{moment} partitioning, but the results are equally applicable to force partitioning. For simplicity, we continue to refer to the analysis technique by its full name (or acronym) -- FMPM.

In the following sections, we introduce the experimental setup, the vortex identification method, and the FMPM (Section \ref{sec.methods}), including the relative benefits of applying FMPM to instantaneous (noisy) velocity fields as compared with ensemble-averaged (less noisy) fields. In Section \ref{sec.results}, we analyze and scale the effect of pitching frequency, amplitude and pivot location on the temporal evolution of the vortex trajectory, circulation and the FMPM-based vorticity-induced moment. Finally, we summarize all the key findings and reiterate the significance of the present study in Section \ref{sec.conclusion}.

\section{Methods}\label{sec.methods}
\subsection{Experimental setup}\label{sec.setup}

\begin{figure*}
\centering
\includegraphics[width=1\textwidth]{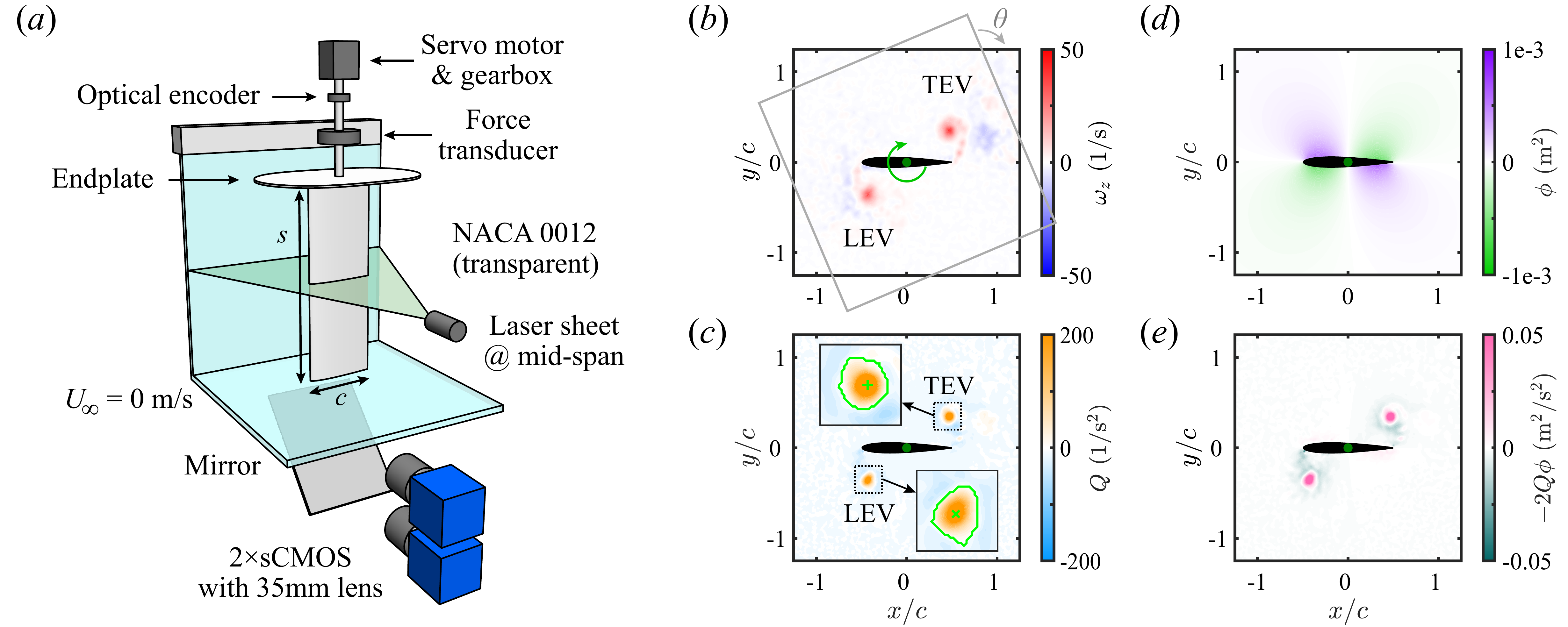}
\caption{(\emph{a}) A schematic of the experimental setup. (\emph{b}) A sample phase-averaged spanwise vorticity field, $\omega_z$, for the case $A=30^\circ$, $f=0.5$ Hz and $x/c=0.5$ during the pitch-up motion ($t/T=0.13$, where $T=1/f$ is the pitching period). The original frame (i.e. the gray box) is rotated by $\theta$ to keep the wing at zero angle-of-attack. (\emph{c}) The corresponding $Q$ field. Insets are zoom-in views of the LEV and TEV, with the identified vortex positions and boundaries. (\emph{d}) The influence field, $\phi$ (i.e. the auxiliary potential). (\emph{e}) The vortex-induced moment density distribution, $-2Q\phi$.}
\label{fig.setup}
\end{figure*}

A schematic of the experimental setup is shown in Fig. \ref{fig.setup}(\emph{a}). The setup is very similar to that used in \citet{zhu2021nonlinear}, but without the cyber-physical control loop. We conduct all the experiments in the Brown University free-surface water tunnel (test section $\mathrm{width} \times \mathrm{depth} \times \mathrm{length} = 0.8~\mathrm{m}\times0.6~\mathrm{m}\times4.0~\mathrm{m}$), with a zero flow speed ($U_{\infty}=0$ m/s) to create a quiescent environment. We use a servo motor (Parker SM233AE) with a 5:1 gearbox to pitch a transparent acrylic NACA 0012 wing (span $s=0.3$ m, chord $c=0.1$ m, aspect ratio $AR=3$). The wing has an endplate on the top end to reduce three-dimensional effects and skim surface waves at the root; the wing tip (bottom end) is left free. The wing is mounted above the endplate (not shown) using an adjustable bracket that allows the pitching axis location $x/c$ to be varied between 0 and 1. In between the servo motor and the wing, an optical encoder (US Digital E3-2500) is used to measure the pitching position $\theta$, and a six-axis force/torque transducer (ATI 9105-TIF-Delta-IP65) is used to measure the fluid forces and torques exerted on the wing. In the experiments, we pitch the wing using a sinusoidal profile $\theta = A \sin{(2 \pi f t)}$, where $A$ is the pitching amplitude, $f$ is the pitching frequency and $t$ is time. In this study, we focus on four pitching amplitudes: $A=30^\circ$ (0.52 rad), $60^\circ$ (1.05 rad), $90^\circ$ (1.57 rad) and $120^\circ$ (2.09 rad), four pitching frequencies: $f=0.25$, 0.5, 0.75 and 1.0 Hz, and three pitching axis locations: $x/c=0.5$ (mid-chord), $x/c=0.25$ (quarter-chord) and $x/c=0$ (leading-edge). Because we pitch the wing sinusoidally, we can approximate the averaged angular velocity as $\dot{\theta}=4Af$ (i.e. the wing rotates four times the pitching amplitude $A$ over one cycle $1/f$) and define the Reynolds number as $Re \equiv 4\rho A f c_m^2/\mu$, where $c_m$ is the effective chord length, $\rho$ and $\mu$ are water density and dynamic viscosity, respectively. $Re$ is of $\mathcal{O}(10^4)$ for the wing kinematics considered in the present study.

To study the vortex dynamics associated with the pitching wing, we use a two-dimensional PIV system to measure the flow field around the wing. We seed the water with 50 $\mathrm{\mu m}$ diameter neutrally buoyant hollow ceramic spheres and illuminate the mid-span plane using a double-pulse Nd:YAG laser (532 nm, Quantel EverGreen) with LaVision sheet optics. Two co-planar sCMOS cameras (LaVision, 2560 $\times$ 2160 pixels) with a $45^\circ$ mirror are used to record PIV image pairs at a frame rate of 15 Hz. The recorded PIV images are processed using the DaVis software (v10, LaVision, two passes at $64\times64$ pixels, two passes at $32\times32$ pixels, both with 50\% overlap) to calculate the velocity vectors. The DaVis software estimates the uncertainty in velocity vectors to be around 5\% of the maximum velocity. Finally, the velocity fields obtained from the two cameras are stitched together to form a $3.2c \times 3.2c$ field of view.

The laser and cameras are phase-synchronized with the pitching motion for phase-averaging the velocity field. The laser/camera frequency is maintained at 15 Hz, and the number of frames (bins) per pitching cycle is determined by the pitching frequency. For example, a pitching frequency of $f=0.5$ Hz will result in 30 frames per cycle. All velocity fields are phase-averaged over 20 cycles. This means 600 instantaneously measured velocity fields are needed for phase-averaging a $f=0.5$ Hz case. The standard deviation of the velocity vectors within a phase is around 10\% of the average velocity.

The spanwise vorticity field, $\omega_z=\partial v/\partial x - \partial u/\partial y$, was calculated using the DaVis software based on the central difference scheme with the four closest neighbors. A sample spanwise vorticity field for the case $A=30^\circ$, $f=0.5$ Hz and $x/c=0.5$ during the pitch-up motion is shown in Fig. \ref{fig.setup}(\emph{b}). For visualization purposes, we rotate the original frame (i.e. the gray box) by the pitching angle, $\theta$, to keep the wing at zero angle-of-attack. We see that as the wing pitches up, two patches of positive spanwise vorticity are generated at the leading edge and the trailing edge, corresponding to the leading-edge vortex (LEV) and the trailing-edge vortex (TEV), respectively. The LEV/TEV shear layers are also visible, although they start to break into smaller secondary vortices \citep{francescangeli2021discrete}. Two patches of negative vorticity are seen near the positive vortices, corresponding to the negative LEV and TEV left over from the previous pitch-down cycle.

\subsection{Vortex identification and trajectory tracking}\label{sec.vortex_id}

We identify the vortex cores and boundaries using the $Q$-criterion \citep{hunt1988eddies,jeong1995identification}, 
\begin{equation}
Q=\frac{1}{2}(\Vert\boldsymbol{\Omega}\Vert^2-\Vert\boldsymbol{\mathrm{S}}\Vert^2), 
\label{eq:Q-def}
\end{equation}
where $Q$ is the second invariant of the velocity gradient tensor, $\boldsymbol{\Omega}$ is the vorticity tensor and $\boldsymbol{\mathrm{S}}$ is the strain-rate tensor. Connected regions with $Q>0$, where rotation is higher than strain, are identified as a vortex \citep{lee2022leading}. The position of the vortex core is calculated as the centroid of the top ten $Q$ values within the vortex boundary. The $Q$ field corresponding to Fig. \ref{fig.setup}(\emph{b}) is plotted in Fig. \ref{fig.setup}(\emph{c}). The insets are zoom-in views of the LEV and TEV, with the identified vortex positions and boundaries. We see that the vortex boundaries ($Q=0$, green curves) faithfully capture the LEV and TEV domains. There are regions with negative $Q$ values surrounding the vortices, corresponding to strain-dominated regions. These strain-dominated regions have been shown to contribute to the generation of opposite-signed aerodynamic loads \citep{menon2021significance}. The spanwise circulation of the vortex, $\Gamma$, is evaluated by integrating the spanwise vorticity, $\omega_z$, within the vortex boundary using Stokes' theorem. Note that calculating the circulation from the phase-averaged velocity field is equivalent to taking the phase average of individual circulations, as calculating the vortex circulation is a linear operation.

\subsection{Force and moment partitioning method (FMPM)}\label{sec.MPM}

For convenience and completeness, we first review the FMPM approach. Following this, we will consider the extension of the method to ensemble-averaged velocity fields. Following \citet{menon2021quantitative}, we first construct an auxiliary potential, $\phi$, with
\begin{equation}
    \nabla^2 \phi = 0,~\frac{\partial \phi}{\partial \boldsymbol{\mathrm{n}}} =
    \begin{cases}
      [(\boldsymbol{x}-\boldsymbol{x_p})\times\boldsymbol{\mathrm{n}}]\cdot\boldsymbol{\mathrm{e_z}}~~\text{on airfoil} \\
      0~~\text{on outer boundary}
    \end{cases},
\label{eqn.phi}
\end{equation}
where $\boldsymbol{\mathrm{n}}$ is the outward-facing unit vector normal to the boundary, $\boldsymbol{x}-\boldsymbol{x_p}$ is the location vector pointing from the pitching axis $\boldsymbol{x_p}$ towards the location $\boldsymbol{x}$ on the surface of the airfoil, and $\boldsymbol{\mathrm{e_z}}$ is the spanwise unit vector. Note that this auxiliary potential $\phi$ is specifically constructed for \emph{moment} partitioning. A different potential can be constructed to determine lift forces or drag forces, etc \citep{menon2021quantitative,menon2021significance}. In addition, the FMPM potential, which we refer to as the ``influence field'', should not be confused with the more familiar velocity potential from the classical irrotational flow theory. The influence field quantifies the spatial influence of the $Q$-field on the resultant moment acting on the submerged body.

The influence field, $\phi$, satisfies the Laplace equation, subject to two different Neumann boundary conditions on the airfoil and the outer boundary. It is only a function of the airfoil shape, its orientation, and the location of the pitching axis. We solve Eqn. \ref{eqn.phi} numerically using the MATLAB Partial Differential Equation Toolbox (Finite Element Method, code available on \href{https://www.mathworks.com/matlabcentral/fileexchange/130274-force-moment-partitioning-influence-potential-solver-2d}{\textcolor{blue}{MATLAB File Exchange}}). Fig. \ref{fig.setup}(\emph{d}) shows the calculated influence field, $\phi$, for a NACA 0012 airfoil pitching at the mid-chord. For this choice of the pitching axis, we see that the $\phi$ field can be divided into four quadrants, with the upper surface of the fore wing and the lower surface of the aft wing being positive, and the upper surface of the aft wing and the lower surface of the fore wing being negative. The magnitude of $\phi$ is the highest near the wing surface and decreases with distance. The $\phi$ field is not exactly symmetric about the mid-chord due to the airfoil shape, with the zero-$\phi$ boundary slightly shifted towards the trailing edge.

The vortex-induced force/moment \emph{density} is expressed in terms of $Q$ \citep{menon2021quantitative} as follows:
\begin{equation}
     f_Q = -2 \rho Q \phi,
\end{equation}
where $\rho$ is the fluid density. The vortex-induced force/moment is then given by the integral of $f_Q$: 
\begin{equation}
    \tau = -2\rho \int_V Q \phi~\mathrm{d}V,
\label{eqn.vortex_torque}
\end{equation}
where $\int_V$ represents volume integral over the field of interest. As mentioned above, we focus on the torque, $\tau$, in this paper. However, as shown in \citet{menon2021initiation,menon2021significance}, with the appropriate influence field, any force can be computed in this manner.

The spatial distribution of the vorticity-induced torque (moment) near the pitching wing can thus be visualized by plotting contours of $-2Q\phi$ (i.e. the vorticity-induced moment density distribution). Fig. \ref{fig.setup}(\emph{e}) shows the moment density distribution corresponding to Fig. \ref{fig.setup}(\emph{b}). As expected, in this small-amplitude pitching case ($A=30^\circ$), the LEV and TEV both contribute to positive (counterclockwise) moments. However, as we will show later, the vortex-induced moment can switch signs during the pitch-up/down cycle for large pitching amplitudes.

The Force and Moment Partitioning Method can also be used to obtain the added-mass torque \citep{menon2021initiation,menon2021quantitative,menon2021significance}
\begin{equation}
    \tau_a = -\rho \int_S \boldsymbol{\mathrm{n}} \cdot \Bigl(\frac{\mathrm{d}\boldsymbol{U}}{\mathrm{d}t} \phi \Bigl)~\mathrm{d}S,
\label{eqn.added_mass}
\end{equation}
where $\boldsymbol{U}$ is the velocity of the moving wing boundary, $\phi$ is the same influence field calculated using Eqn. \ref{eqn.phi}, and $\int_S$ is the integral along the wing surface. We see from Eqn. \ref{eqn.added_mass} that the added-mass torque is only a function of the wing geometry and kinematics, flow field data is not required for calculating $\tau_a$. We want to note that there have been recent developments in using PIV data to quantify the added-mass forces \citep{corkery2019quantification,gehlert2021noncirculatory}. In \citet[][see their Eqn. 5 and the associated section]{zhang2015centripetal}, this term was referred to as the ``centripetal acceleration reaction'' (as compared to the conventional added mass, which is a linear acceleration reaction term), and it only exists when the no-slip boundary condition is satisfied. Therefore, in the context of FMPM, this term is included in the added-mass term (Eqn. \ref{eqn.added_mass}) and not the vorticity-induced term (Eqn. \ref{eqn.vortex_torque}), which does not include the vorticity on the boundary.

The torque due to viscous diffusion can also be calculated using FMPM \citep{menon2021quantitative}. However, it is negligible for the present study due to the relatively high Reynolds number $Re\sim\mathcal{O}(10^4)$.

\subsection{Application of FMPM to ensemble- and phase-averaged data}\label{sec.phase_avg}

In many experiments, including those discussed in this paper, measured velocity data may be ensemble-, or phase-averaged so as to remove noise or turbulence-related flow structures. In what follows, we consider ensemble- and phase-averaged data as interchangeable, both representing time-dependent averages of a more complex and noisy flow. Application of FMPM to phase-averaged data introduces some subtleties because $Q$ is a nonlinear function of velocity (Eqn. \ref{eq:Q-def}). In the most general formulation, any quantity, for example the velocity $u$ that is measured in an experiment, can be expressed as a sum of the time-averaged value, $\bar{u}$, the phase-averaged value, $\langle u \rangle$, and the instantaneous fluctuations, $u'$:
\begin{equation}
    u(x,y,z,t) = \bar{u}(x,y,z) + \langle u \rangle (x,y,z,t) + u'(x,y,z,t).
\label{eqn.u_decom}
\end{equation}
The quantity $Q$ can be expressed as
\begin{equation}
    Q = -0.5\frac{\partial}{\partial x_i}(u_j\frac{\partial u_i}{\partial x_j}) \equiv -0.5u_{i,j}u_{j,i}, 
\end{equation}
where indicial notation, with implied summation, has been used. 
As with the velocity, $Q$ can be expressed as
\begin{equation}
    Q(x,y,z,t) = \bar{Q}(x,y,z) + \langle{Q}\rangle(x,y,z,t) + Q'(x,y,z,t).
\end{equation}
We can obtain the expression for the above $Q$ components in terms of velocity as follows
\begin{equation}
    Q = Q_{\bar{u}\bar{u}} + Q_{\langle{u}\rangle\langle{u}\rangle} + Q_{u' u'} + 2Q_{\bar{u}\langle{u}\rangle} + 2Q_{\bar{u} u'} + 2Q_{\langle{u}\rangle u'},
\end{equation}
where $Q_{\bar{u}\langle{u}\rangle}=-0.5\bar{u}_{i,j}\langle{u}_{j,i}\rangle$, etc. Taking a time average of the above expression, we can get $\bar{Q} \equiv Q_{\bar{u}\bar{u}}$.
Taking a phase average of the above expression gives
\begin{equation}
\begin{split}
    \langle{Q}\rangle = &~\cancel{\langle{Q}_{\bar{u}\bar{u}}\rangle} + \langle{Q}_{\langle{u}\rangle\langle{u}\rangle}\rangle + \langle{Q}_{u'u'}\rangle \\
     &~~~~~~~~~+ 2\langle{Q}_{\bar{u}\langle{u}\rangle}\rangle + \cancel{2\langle{Q}_{\bar{u} u'}\rangle} + 2\langle{Q}_{\langle{u}\rangle u'}\rangle \\
    \equiv &~Q_{\langle{u}\rangle\langle{u}\rangle} + 2Q_{\bar{u}\langle{u}\rangle} + 2\langle{Q}_{\langle{u}\rangle u'}\rangle + \langle{Q}_{u' u'}\rangle.
\end{split}
\end{equation}
The phase-averaged vortex-induced force/moment density is given by
\begin{equation}
\begin{split}
    & \langle{f}_Q\rangle(x,y,z,t) = -2\rho\langle{Q}\rangle\phi \\
    & = -2\rho[Q_{\langle{u}\rangle\langle{u}\rangle} + 2Q_{\bar{u}\langle{u}\rangle} + 2\langle{Q}_{\langle{u}\rangle u'}\rangle + \langle{Q}_{u' u'}\rangle]\phi.
\end{split}
\end{equation}
If the flow is highly turbulent and the time-dependent fluctuations are large, the last two terms in the above equation should not be ignored. However, for laminar flows, where we can assume $\vert u' \vert << \vert \langle{u}\rangle \vert$, we have
\begin{equation}
\begin{split}
    \langle{f}_Q\rangle(x,y,z,t) & \approx -2\rho\langle{Q}\rangle\phi \\
    & = -2\rho[Q_{\langle{u}\rangle\langle{u}\rangle} + 2Q_{\bar{u}\langle{u}\rangle}]\phi.
\end{split}
\label{eqn.phase_avg}
\end{equation}
This assumption will be tested for the present data in the next section. Thus, to obtain an accurate estimation of $\langle{f}_Q\rangle$ from velocity data, we should either compute $\langle{Q}\rangle$ from the instantaneous $Q$ field directly (i.e. calculate $Q$ from the instantaneous velocity field and then phase average), or estimate the sum of $Q_{\langle{u}\rangle\langle{u}\rangle} + 2Q_{\bar{u}\langle{u}\rangle}$. The estimation of $\bar{u}$ is not straightforward in a flow with moving boundaries such as oscillating foils, since it is not clear how to obtain time-averaged data in regions through which the foil passes over time. One possible approach is simply to set $\bar{u}$ equal to zero and only consider the phased-averaged flow field, $\langle u \rangle$.  An alternative way of calculating $\bar{u}$ is to use Dynamic Mode Decomposition (DMD) \citep{menon2020dynamic}, since the zeroth DMD mode is the time-averaged mean. However, as we will show later, for the present study where there is no free-stream flow, the $2Q_{\bar{u}\langle{u}\rangle}$ term can be safely neglected. Our analysis in this subsection (Eqns. \ref{eqn.u_decom} - \ref{eqn.phase_avg}) outlines the potential subtleties for applying FMPM to ensemble- or phase-averaged data. We want to emphasize that our analysis is \emph{general}, and additional tweaks might be needed for adapting the analysis to specific situations.

\section{Results and Discussion}{\label{sec.results}}
\subsection{Scaling of vortex circulation and trajectory}{\label{sec.traj_circ}}

\begin{figure*}
\centering
\includegraphics[width=0.9\textwidth]{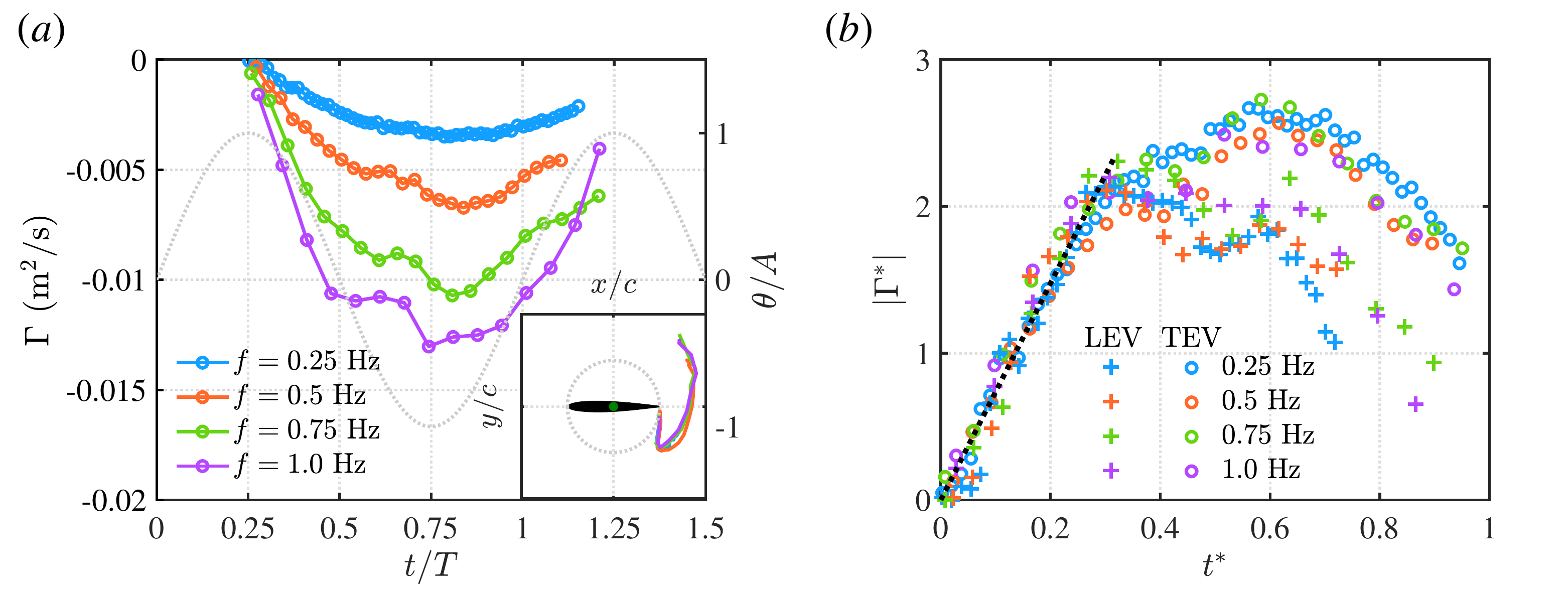}
\caption{(\emph{a}) Trailing-edge vortex (TEV) circulation, $\Gamma$, during pitch-down for $A=30^\circ$, $x/c=0.5$ and $f=0.25$ to 1.0 Hz. The gray dotted curve represents the non-dimensional pitching position, $\theta/A$, on the right axis. Inset: TEV trajectories for the corresponding cases. (\emph{b}) Magnitudes of the non-dimensional circulations, $\vert\Gamma^*\vert$ (Eqn. \ref{eqn.gamma_star}), for both LEVs (plus signs) and TEVs (circles), with the marker colors corresponding to the frequencies in (\emph{a}). The black dotted line shows the linear trend for data points within the initial linear growth regime, $0 \leq t^* \leq 0.3$.}
\label{fig.circ_f_scale}
\end{figure*}

To characterize the vortex dynamics associated with the pitching wing in quiescent water, we first analyze the vortex trajectories and circulation. Because we pitch the wing sinusoidally, the wing moves symmetrically during pitch-up and pitch-down motions. Moreover, \citet{zhu2021nonlinear} have shown that the fluid damping induced by LEVs and TEVs are comparable, although there are subtle differences
caused by the rounded leading edge and the sharp trailing edge of the NACA 0012 airfoil. Therefore, in this study, we choose to focus on analyzing the \emph{TEV} dynamics during the \emph{pitch-down} motion. The LEV dynamics will only be analyzed for some cases for comparison.

Fig. \ref{fig.circ_f_scale}(\emph{a}) shows the TEV circulation, $\Gamma$, during pitch-down for a wing pitching at the mid-chord, $x/c=0.5$, with a pitching amplitude of $A=30^\circ$, and a pitching frequency varied from $f=0.25$ to 1.0 Hz. The onset of TEV growth starts at $t/T = 0.25$ for all pitching frequencies because we pitch the wing \emph{sinusoidally}. At $t/T = 0.25$, the pitching velocity starts to increase from zero as the pitch reversal starts, and the TEV starts to emerge and grow. Because the wing undergoes a pitch-down motion, the TEV has negative circulation. We see that for a fixed pitching frequency, the TEV circulation starts from zero and decreases as the wing pitches downward, as more negative vorticity is fed into the TEV through the shear layer. This process continues until $t/T=0.75$, when the pitch-up motion starts and the connection between the TEV and the shear layer is cut off. A bump in $\Gamma$ for $f=0.75$ and $f=1.0$ Hz shows up right before $t/T=0.75$ when the TEV starts to separate from the shear layer. After $t/T\approx0.75$, the TEV begins to decline as there is no new vorticity input and the existing vortex starts to dissipate. The magnitude of the TEV circulation increases with the pitching frequency, as the vortex feeding shear-layer velocity increases linearly with the pitching frequency \citep{onoue2016vortex,onoue2017scaling}.

The inset of Fig. \ref{fig.circ_f_scale}(\emph{a}) shows the TEV trajectories for the corresponding four pitching frequencies. The figure frame is rotated so as to keep the wing at zero pitching angle. It is observed that for a fixed pitching amplitude, TEV trajectories collapse well for different pitching frequencies. A circle centered at the mid-chord with a diameter $c$ is plotted in gray dots to illustrate the trailing edge trajectory. The initial part of the TEV trajectory is almost perpendicular to the wing chord, confirming the validity of the linear assumptions used in the fluid damping scaling proposed in \citet{zhu2021nonlinear} for small pitching amplitudes. As the pitch reversal starts, the TEV trajectories turn abruptly upwards. We see that the TEV trajectory deviates from the trailing edge trajectory. This is in contrast to the results of \citet{francescangeli2021discrete}, who observed that the shed vortex closely follows the trajectory of the edge of a flat plate pitching with a trapezoidal velocity profile. In our sinusoidal pitching case, the deviation between the TEV trajectory and the trailing edge trajectory presumably comes from two effects: the interaction between the TEV and the opposite-signed residual vortex from the previous pitch-up motion, and the weak ambient flow induced by the sinusoidal pitching motion \citep{shinde2013jet}.

\citet{onoue2016vortex} showed that for a pitching plate undergoing large-amplitude limit-cycle oscillations in a freestream flow, the LEV circulation scales with the feeding shear-layer velocity multiplied by a characteristic length scale. Following a similar approach, we propose an LEV/TEV circulation scaling for pitching wings in quiescent flow. Because the freestream velocity is zero, the feeding shear-layer velocity equals the leading-edge/trailing-edge velocity, which is given by $U_{SL}=4Afc_{m}$, where $c_m$ represents the distance between the leading/trailing edge and the pivot point (i.e. the effective chord length). Therefore, we can write the non-dimensional circulation as
\begin{equation}
      \Gamma^* = \frac{\Gamma}{4Afc_m^2}.
\label{eqn.gamma_star}
\end{equation}
We note that the definition of $\Gamma^*$ is analogous to the vortex formation number, $\hat{T}$, which quantifies the growth of a vortex, and the maximum value (i.e. the optimal vortex formation number) represents when the vortex stops entraining additional vorticities from the feeding shear layer. We can also non-dimensionalize the vortex formation time as
\begin{equation}
      t^* = \frac{4 A c_{m}}{c}(t/T-0.25),
\label{eqn.t_star}
\end{equation}
where $T=1/f$ is the pitching period; we use the $t/T-0.25$ term to offset the starting time to coincide with the start of LEV/TEV formation. In $\Gamma^*$ and $t^*$, $c_m = c_{LE}$ for LEVs and $c_m = c_{TE}$ for TEVs, where $c_{LE}$ and $c_{TE}$ represent the leading-edge chord and the trailing-edge chord, respectively, and $c=c_{LE}+c_{TE}$ is the full chord length of the wing.

Fig. \ref{fig.circ_f_scale}(\emph{b}) shows the evolution of the magnitude of the TEV circulation, $\vert\Gamma^*\vert$ (circles), as a function of the vortex formation time, $t^*$, for four different pitching frequencies ($f=0.25$, 0.5, 0.75 and 1.0 Hz), corresponding to the cases plotted in Fig. \ref{fig.circ_f_scale}(\emph{a}). The scaled LEV circulations at these frequencies are also plotted using plus signs for comparison. We see that the data points collapse well under the proposed scaling, showing the frequency dependence of the circulation. An initial linear growth regime ($0 \leq t^* \leq 0.3$) is observed for both the LEV and the TEV circulations. In this linear regime, LEV and TEV circulations overlap and no significant difference in their slopes is observed. After the linear regime, the TEV circulation (circles) keeps on increasing and reaches its maximum $\vert\Gamma^*\vert\approx 3$ at $t^* \approx 0.6$. On the other hand, the LEV circulation (plus signs) decreases right after the linear regime despite that the pitch reversal starts around $t^*\approx0.5$. This difference is believed to result from the fact that the TEVs generated by the sharp trailing edge are more coherent so that they can sustain longer than the LEVs generated by the rounded leading edge. This difference is clearly captured by the vorticity field shown in Fig. \ref{fig.setup}(\emph{b}). We see that at this time instant ($t^*\approx0.9$), the negative vorticities from the pitch-down motion (i.e. the blue regions) still retain a circular shape for the TEV, but become unidentifiable for the LEV. The difference in the circulation also echoes the observations of \citet{zhu2021nonlinear} that the fluid damping associated with a sharp trailing edge is higher than that resulted from a rounded leading edge.

\begin{figure*}
\centering
\includegraphics[width=0.9\textwidth]{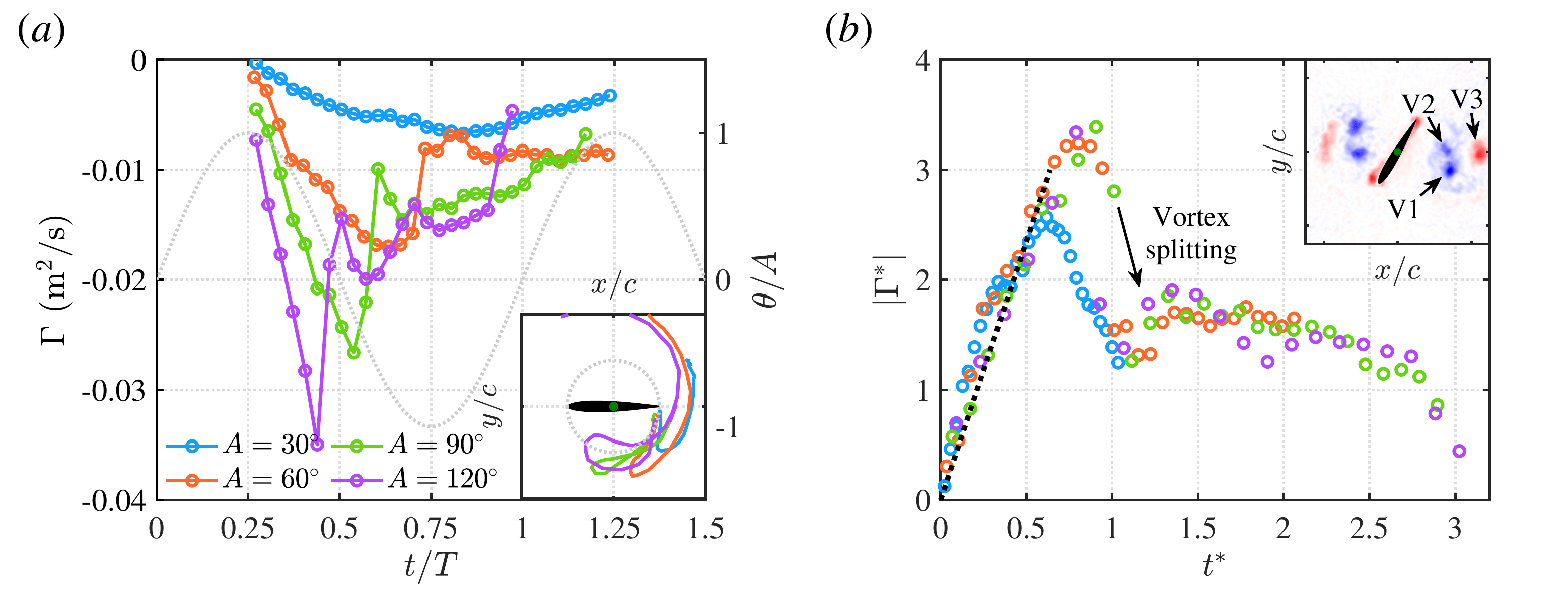}
\caption{(\emph{a}) TEV circulation, $\Gamma$, during pitch-down for $f=0.5$ Hz, $x/c=0.5$ and $A=30^\circ$ to $120^\circ$, with the non-dimensional pitching position, $\theta/A$, plotted on the right axis. Inset: TEV trajectories for the corresponding cases. (\emph{b}) Magnitudes of the non-dimensional TEV circulation, $\vert\Gamma^*\vert$. The black dotted line shows the linear trend for data points within the initial linear growth regime, $0 \leq t^* \leq 0.6$. The inset is a sample spanwise vorticity field for $A=60^\circ$ at $t/T\approx0.75$. At this time instant, the TEV (as well as the LEV) split into two smaller vortices, V1 and V2, with V1 persisting and V2 quickly dissipating away. The opposite-signed vortex V3 is the residual TEV from the previous pitch-up motion.}
\label{fig.circ_amp_scale}
\end{figure*}

Next, we look into the effect of pitching amplitude on the TEV circulation and trajectory. Fig. \ref{fig.circ_amp_scale}(\emph{a}) shows the TEV circulation, $\Gamma$, during pitch-down for a wing pitching around the mid-chord $x/c=0.5$ and at a frequency of $f=0.5$ Hz, with the pitching amplitude varied from $A=30^\circ$ to $120^\circ$. We see that, again, the TEV circulation decreases from zero when the pitch-down motion starts at $t/T=0.25$. The magnitude of $\Gamma$ increases with $A$ due to the higher feeding shear-layer velocity induced at higher pitching amplitudes. There exists a linear growth regime for $\vert\Gamma\vert$ at all four pitching amplitudes, and this regime shrinks as $A$ increases. Following this regime, we observe an abrupt drop of the TEV circulation magnitude for $A=60^\circ$ to $120^\circ$, which we attribute to vortex splitting. This vortex splitting behavior is depicted in Fig. \ref{fig.circ_amp_scale}(\emph{b}) inset, where we plot a sample spanwise vorticity field for $A=60^\circ$ at $t/T\approx0.75$. At this time instant, the TEV (as well as the LEV) split into two smaller vortices, V1 and V2, with V1 persisting and V2 quickly dissipating away. As such, the circulation for V1 is tracked after this split. After the split, some surrounding vorticities are re-entrained into V1, resulting in a slight increase in the circulation magnitude right after the drop. The opposite-signed vortex V3 is the residual TEV from the previous pitch-up motion. No TEV splitting is observed for the smallest pitching amplitude $A=30^\circ$. The origin of this vortex splitting behavior is unclear but worth further investigation.

Fig. \ref{fig.circ_amp_scale}(\emph{a}) inset shows the TEV trajectories for different pitching amplitudes. We see that as the pitching amplitude increases, the TEV trajectory no longer follows the perpendicular path observed for the lowest pitching amplitude case (i.e. $A=30^\circ$, the blue curve, see also Fig. \ref{fig.circ_f_scale}\emph{a} inset). Instead, the TEV starts to loosely follow the trailing edge trajectory (i.e. the gray dotted circle) at the beginning of the pitch-down motion. As the pitching amplitude further increases ($A=90^\circ$ and $120^\circ$), the TEV trajectory sees the emergence of a turn-over loop -- the TEV moves to the front of the pivot axis, turns closer to the wing surface, intersects with its initial path and eventually dissipates to the other side of the wing. The vortex trajectories cannot be simply scaled by the pitching amplitude because of their complex geometries. The vortex splitting and the complex vortex trajectories both add complexities to the problem, causing the nonlinear behaviors of the fluid damping at higher pitching amplitudes observed in \citet{zhu2021nonlinear}. A more quantitative analysis of this issue and the connections between vortex dynamics and fluid damping will be discussed later in \ref{sec.FMPM}.

We scale the TEV circulations as well as the vortex formation time for different pitching amplitudes using Eqns. \ref{eqn.gamma_star} and \ref{eqn.t_star}, respectively, and show the results in Fig. \ref{fig.circ_amp_scale}(\emph{b}). Once again, the TEV circulation collapses well under the proposed $\Gamma^*$ scaling, revealing the amplitude dependence of the vortex strength. The $A$ term in the $t^*$ scaling (Eqn. \ref{eqn.t_star}) aligns the location of the maximum $\Gamma$ for different pitching amplitudes. The initial linear growth regime for moderate to large pitching amplitudes ($A=60^\circ$ to $120^\circ$) extends to $t^*>0.6$ as compared to $A=30^\circ$ (see also Fig. \ref{fig.circ_f_scale}\emph{b}), with the maximum $\vert\Gamma^*\vert$ elevated to $\vert\Gamma^*\vert>3$. After the linear regime and the abrupt amplitude drop caused by vortex splitting, the scaled vortex circulation for $A=60^\circ$ to $120^\circ$ decays slowly with a relatively constant slope at $t^*>1.2$ \citep[i.e. vortex saturation,][]{devoria2012vortex}. The $\vert\Gamma^*\vert$ for $A=30^\circ$ does not reach this slow-decay regime before the vortex boundary becomes unidentifiable.

\begin{figure*}
\centering
\includegraphics[width=0.9\textwidth]{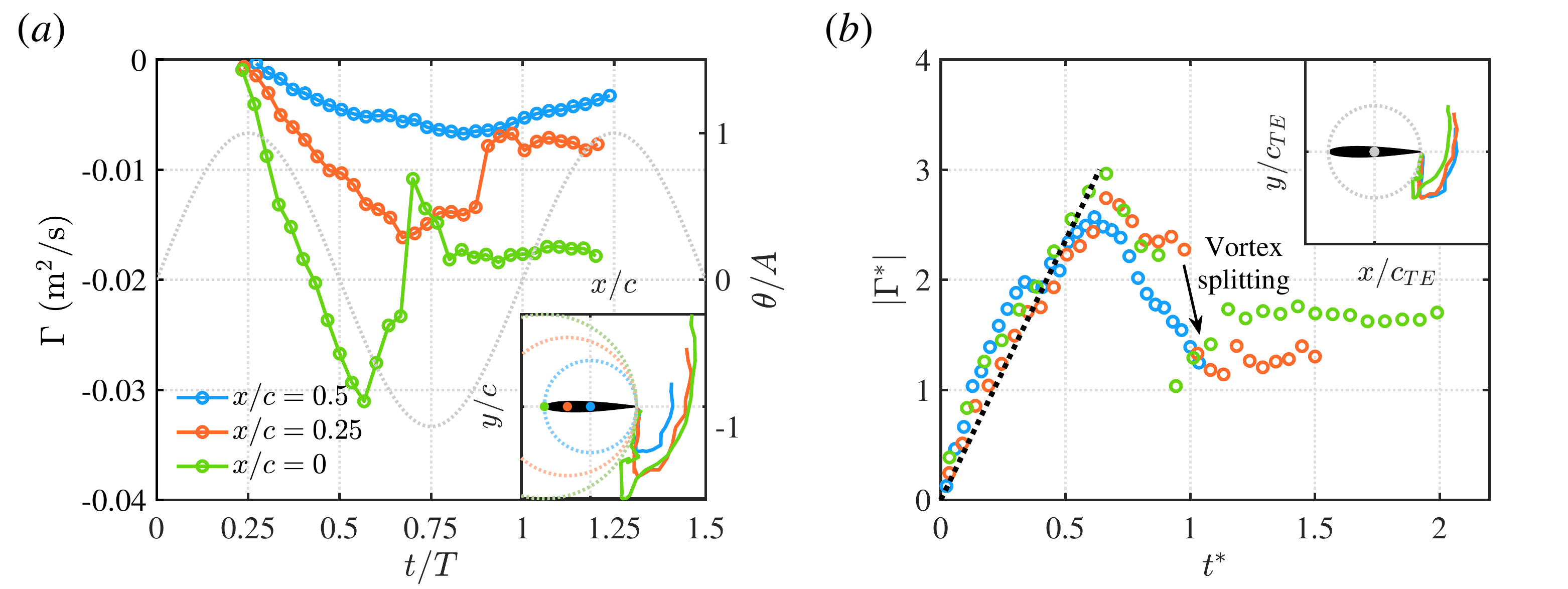}
\caption{(\emph{a}) TEV circulation, $\Gamma$, during pitch-down for $A=30^\circ$, $f=0.5$ Hz, $x/c=0.5$ (mid-chord pitching), $x/c=0.25$ (quarter-chord pitching) and $x/c=0$ (leading-edge pitching), with the pitching position, $\theta/A$, on the right axis. Inset: TEV trajectories for the corresponding cases. (\emph{b}) Magnitudes of the non-dimensional TEV circulation, $\vert\Gamma^*\vert$, with a black dotted line showing the linear trend for data points within the initial linear growth regime, $0 \leq t^* \leq 0.6$. Inset: TEV trajectory scaling.}
\label{fig.circ_axis_scale}
\end{figure*}

Another important parameter governing the LEV/TEV dynamics is the location of the pitching axis. Fig. \ref{fig.circ_axis_scale}(\emph{a}) shows the temporal evolution of the TEV circulation, $\Gamma$, during the pitch-down motion for a wing pitching at a frequency of $f=0.5$ Hz and an amplitude of $A=30^\circ$, with the pitching axis located at the mid-chord ($x/c=0.5$), the quarter-chord ($x/c=0.25$) and the leading-edge ($x/c=0$). Similar to previous results for different pitching frequencies and amplitudes, the TEV circulation starts from zero and decreases as the wing pitches down. The circulation magnitude increases as the pitching axis moves farther from the mid-chord, due to the higher feeding shear-layer velocity $U_{SL}=4Afc_{m}$. Again, $\Gamma$ decreases linearly in the early stage of the pitch-down motion, and this linear regime shortens in time as the pitching axis moves towards the leading edge. After the linear regime, the TEV circulation drops near-linearly first and then abruptly for $x/c=0.25$ and $x/c=0$ as the TEV splits into two. The TEV circulation and formation time are scaled using Eqn. \ref{eqn.gamma_star} and \ref{eqn.t_star}, and replotted in Fig. \ref{fig.circ_axis_scale}(\emph{b}). We continue to see a very nice collapse of all the data points under the $\Gamma^*$ scaling, confirming the pitching axis dependence of the vortex strength. The $\vert\Gamma^*\vert$ peaks around $t^*=0.6$, similar to the results observed in Fig. \ref{fig.circ_f_scale}(\emph{b}) and Fig. \ref{fig.circ_amp_scale}(\emph{b}). 

Fig. \ref{fig.circ_axis_scale}(\emph{a}) inset shows the pitching axis locations and the corresponding trailing edge (dotted circles) and trailing-edge vortex (solid curves) trajectories. We see that the three TEV trajectories overlap initially. As the pitching axis moves away from the mid-chord while the angular pitching amplitude is maintained, the curvature of the TE trajectory decreases but its arc length increases. In the meantime, the TEV trajectory scales up in both $x$- and $y$-directions. We then scale both the TE and the TEV trajectories using the trailing-edge chord (i.e. $c_{TE}$, the chord length between the pitching axis and the trailing edge) and the results are shown in the inset of Fig. \ref{fig.circ_axis_scale}(\emph{b}). The three vortex trajectories collapse well under this scaling despite some discrepancies for $x/c=0$ (green curve). This indicates that, unlike the pitching amplitude, which has a nonlinear effect on the vortex trajectory, the location of the pitching axis changes the vortex trajectory in a linear manner. We believe this linear dependence comes from the fact that the trailing edge trajectory (curvature and arc length) is linearly scalable by $c_{TE}$ in the (\emph{x}, \emph{y})-coordinate. On the other hand, the trailing edge trajectories under different pitching amplitudes cannot be simply scaled by $A$ in the (\emph{x}, \emph{y})-coordinate, resulting in the nonlinear unscalable trajectories of the vortex (see Fig. \ref{fig.circ_amp_scale}\emph{a} inset). This argument is also supported by the frequency independence of the TEV trajectories observed in Fig. \ref{fig.circ_f_scale}(\emph{a}) inset, where the trailing edge trajectories remain the same at different pitching frequencies. These results also imply that the vortex trajectory is largely determined by the trailing (or leading) edge trajectory, instead of the edge velocity.

It is also worth noting that in Figs. \ref{fig.circ_f_scale}(\emph{b}), \ref{fig.circ_amp_scale}(\emph{b}) and \ref{fig.circ_axis_scale}(\emph{b}), the maximum $\vert\Gamma^*\vert$ corresponds to the optimal vortex formation number \citep{gharib1998universal,dabiri2009optimal}. However, $\vert\Gamma^*\vert_{max}$ observed in the present study is within a range of 2.5 - 3.5, which is smaller than that observed in previous studies \citep{milano2005uncovering,ringuette2007role,rival2009influence,onoue2016vortex}, where $\hat{T}\approx4$. Because there is no convective free-stream flow in the present study, the vortex formation is dominated by the pitching kinematics (Eqn. \ref{eqn.gamma_star}). Therefore, the vortex is forced to pinch off from the feeding shear layer by the wing kinematics before the universal optimal vortex formation number $\hat{T}\approx4$ can be achieved.

To summarize, the vortex trajectory tracking and scaling analysis performed in this section provides us with many useful insights into the vortex dynamics of pitching wings in a quiescent flow. However, these traditional analysis methods have several limitations. Firstly, vortex tracking and circulation calculation become difficult when the vortex dissipates to an extent that the vortex boundary becomes unidentifiable, and when multiple vortices are in close proximity (e.g. Fig. \ref{fig.circ_amp_scale}\emph{b} inset). In these situations, although the vortices may still contribute to the moment generation, we are not always able to accurately quantify their positions and strengths. Secondly and more importantly, the traditional analysis methods are not capable of directly correlating the spatial position and strength of shed vortices with the resultant vorticity-induced force/moment, which is critical for studying these vortex-dominated flows. Therefore, in the following section, we apply the Force and Moment Partitioning Method to gain more physical insights.

\subsection{Vorticity-induced moment obtained from FMPM}{\label{sec.FMPM}}

\begin{figure*}
\centering
\includegraphics[width=0.9\textwidth]{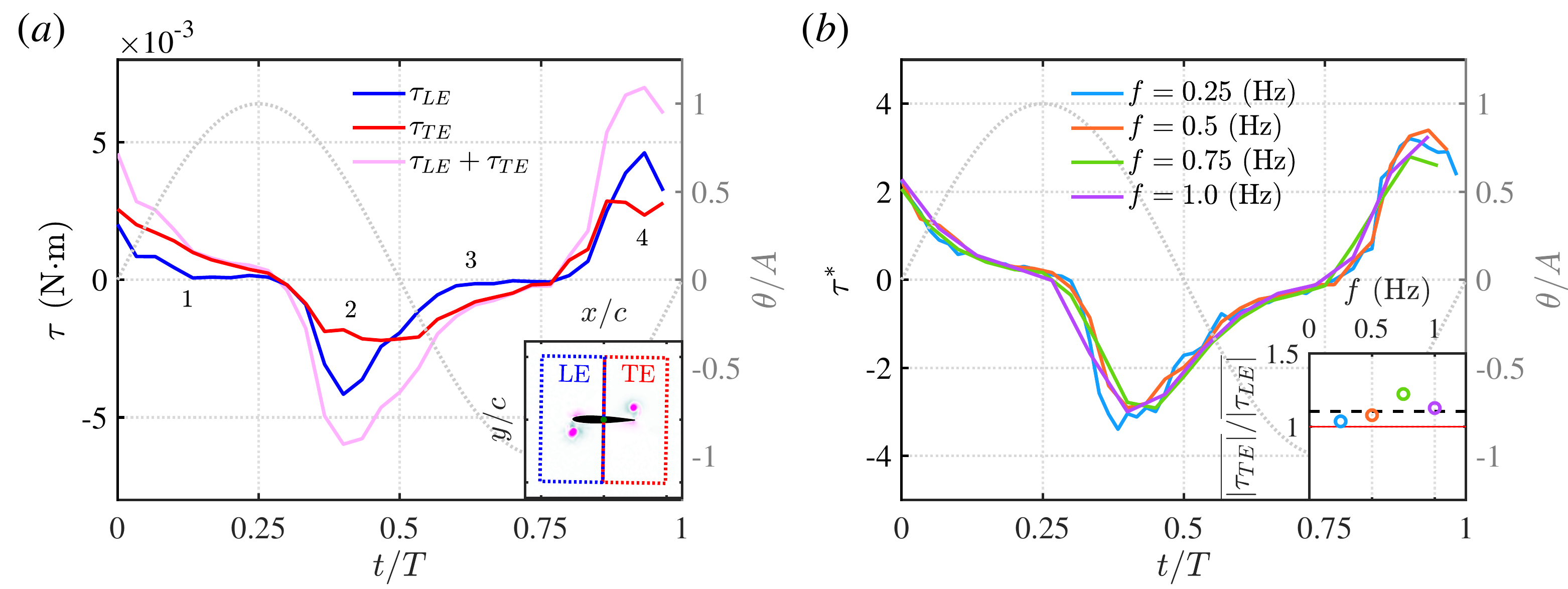}
\caption{(\emph{a}) Time trace of the leading-edge torque ($\tau_{LE}$), trailing-edge torque ($\tau_{TE}$) and total vorticity-induced torque ($\tau_{LE}+\tau_{TE}$) calculated using the PIV-based FMPM for the case $A=30^\circ$, $f=0.5$ Hz and $x/c=0.5$. Inset: Moment density distribution ($-2Q\phi$) at $t/T=0$ and integration windows for the leading-edge and trailing-edge torque. (\emph{b}) Non-dimensional vorticity-induced torque ($\tau^*$) for $A=30^\circ$, $x/c=0.5$ and $f=0.25$ to 1.0 Hz. Inset: Ratio between the cycle-averaged absolute trailing-edge torque and leading-edge torque ($\overline{\vert\tau_{TE}\vert}/\overline{\vert\tau_{LE}}\vert$) for the corresponding frequencies. The black dashed line represents the empirical ratio ($1.05/0.95=1.105$) used in \citet{zhu2021nonlinear}. The red solid line represents a ratio of one.}
\label{fig.torque_LETE_f_scale}
\end{figure*}

The complex behaviors of vortex trajectories and circulations discussed in the above section further affect the corresponding vorticity-induced moment and thus the fluid damping. In this section, we use FMPM to quantify this vorticity-induced torque. The detailed implementation of the FMPM has been introduced in Section \ref{sec.MPM}, and a sample case demonstrating the vorticity field, the $Q$ field, the influence field and the moment density distribution has been shown in Fig. \ref{fig.setup}(\emph{b-e}). The FMPM is not only capable of identifying the total vorticity-induced force/moment, it is also able to separate the force/moment contributions from individual vortices by choosing different integration windows for the expression in Eqn. \ref{eqn.vortex_torque}. Fig. \ref{fig.torque_LETE_f_scale}(\emph{a}) shows the time trace of the leading-edge torque, $\tau_{LE}$, the trailing-edge torque, $\tau_{TE}$, and the total vorticity-induced torque, $\tau=\tau_{LE}+\tau_{TE}$, calculated using the PIV-based FMPM for the case $A=30^\circ$, $f=0.5$ Hz and $x/c=0.5$. The inset shows the integration windows for calculating $\tau_{LE}$ and $\tau_{TE}$. We see that the leading-edge torque and trailing-edge torque behave differently over time, and to analyze, we divide the pitching cycle into four regimes (Fig. \ref{fig.torque_LETE_f_scale}). In regimes 1 and 3, $\tau_{TE}$ has a higher magnitude than $\tau_{LE}$, presumably, as mentioned earlier due to the increased coherence of the TEV as compared to that of the LEV (Fig. \ref{fig.circ_f_scale}\emph{b}). In regimes 2 and 4, where the pitch-down/up motions first start, $\tau_{LE}$ overtakes $\tau_{TE}$ in magnitude because the newly generated LEV stays closer to the wing surface than the TEV. The change of sign in $\tau_{LE}$ and $\tau_{TE}$ aligns with that of the angular pitching velocity, $\dot{\theta}$. 

The total vorticity-induced torque, $\tau=\tau_{LE}+\tau_{TE}$, directly correlates with the fluid damping discussed in \citet{zhu2021nonlinear}, as $\tau = b_f \dot{\theta}$, where $b_f$ is the fluid damping coefficient and $\dot{\theta}$ is the angular velocity. Approximating $\dot{\theta}$ with $4Af$, we modify the fluid damping scaling proposed in \citet{zhu2021nonlinear} to get a vorticity-induced torque scaling for sinusoidally pitching wings in quiescent water
\begin{equation}
    \tau^* = \frac{\tau}{8\rho f^2 A^2 s (K_{LE}c_{LE}^4 + K_{TE}c_{TE}^4)},
\label{eqn.moment_scale}
\end{equation}
where $c_{LE}$ and $c_{TE}$ are the leading-edge chord and trailing-edge chord, respectively. $K_{LE}=0.95$ and $K_{TE}=1.05$ are empirical factors that account for the rounded leading edge and sharp trailing edge, which agree well with experimental observations, as we will show next. 

The frequency-squared dependence of the vorticity-induced torque (Eqn.~\ref{eqn.moment_scale}) is verified in Fig. \ref{fig.torque_LETE_f_scale}(\emph{b}), where we plot the non-dimensional vorticity-induced torque, $\tau^*$, for four different pitching frequencies $f=0.25$, 0.5, 0.75 and 1.0 Hz at a pitching amplitude of $A=30^\circ$ and a pitching axis of $x/c=0.5$. We see that $\tau^*$ collapses nicely under the proposed scaling. Recalling that the vortex trajectory remains unchanged for a wing pitching at a fixed amplitude and different frequencies (Fig. \ref{fig.circ_f_scale}\emph{a}), we know that the weighting by the influence field, $\phi$, also remains unchanged for different pitching frequencies, and we thus conclude that the $\tau \sim f^2$ scaling must come from $Q \sim f^2$ (Eqn. \ref{eqn.vortex_torque}). By definition, $Q$ scales with the vorticity squared, which further scales with circulation (according to Stokes' theorem). Eqn. \ref{eqn.gamma_star} and Fig. \ref{fig.circ_f_scale}(\emph{b}) confirm that $\Gamma \sim f$, and therefore, we infer that $Q \sim f^2$, which again leads to the $\tau \sim f^2$ scaling. These two independent scaling analyses show the self-consistency of our data.

% The leading-edge and trailing-edge torque, $\tau_{LE}$ and $\tau_{TE}$, also individually follow the $\tau^*$ scaling (not shown). 

The inset of Fig. \ref{fig.torque_LETE_f_scale}(\emph{b}) shows the ratio between the cycle-averaged absolute trailing-edge torque and leading-edge torque (i.e. $\overline{\vert\tau_{TE}\vert}/\overline{\vert\tau_{LE}}\vert$) for the corresponding four frequencies, with the black dashed line representing the empirical factors $K_{TE}/K_{LE}=1.05/0.95=1.105$ used in Eqn. \ref{eqn.moment_scale}, and the red solid line representing a ratio of one. We find that the measured ratios (colored circles) are all above one, with a mean value of 1.117, which matches well with the empirical ratio 1.105. This shows that the empirical ratio faithfully captures the subtle differences in the cycle-averaged magnitude of the trailing-edge torque and leading-edge torque, agreeing with the trend observed in \citet{zhu2021nonlinear} that the cycle-averaged trailing-edge fluid damping is slightly higher than that of the leading edge. This also demonstrates that, instead of using an empirical constant, we can also determine the ratio between $K_{LE}$ and $K_{TE}$ using the PIV-based FMPM.

\begin{figure*}
\centering
\includegraphics[width=0.9\textwidth]{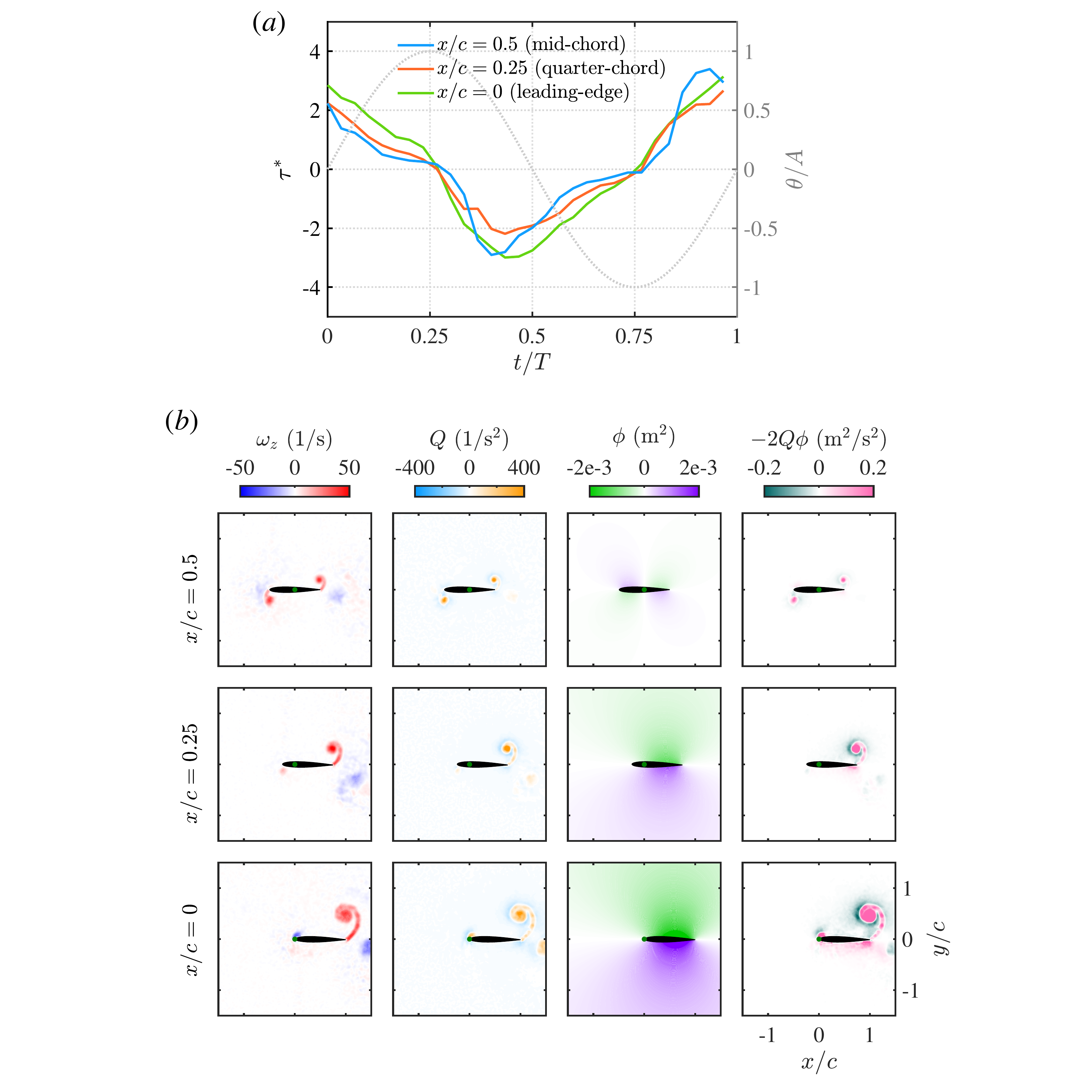}
\caption{(\emph{a}) Non-dimensional vorticity-induced torque ($\tau^*$) for $A=30^\circ$, $f=0.5$ Hz, $x/c=0.5$ (mid-chord pitching), $x/c=0.25$ (quarter-chord pitching) and $x/c=0$ (leading-edge pitching). (\emph{b}) Spanwise vorticity field ($\omega_z$, first column), $Q$ field (second column), influence field ($\phi$, third column) and moment density distribution ($-2Q\phi$, fourth column) for $A=30^\circ$, $f=0.5$ Hz, $x/c=0.5$ (first row), 0.25 (second row) and 0 (third row) at $t/T=0$.}
\label{fig.torque_axis_scale}
\end{figure*}

The scaled torque, Eqn. \ref{eqn.moment_scale}, suggests that the vorticity-induced torque scales with the fourth power of the effective chord length. To validate this, we change the axis of a wing pitching at $A=30^\circ$, $f=0.5$ Hz from $x/c=0.5$ (mid-chord) to $x/c=0.25$ (quarter-chord) and $x/c=0$ (leading-edge) and plot the non-dimensional vorticity-induced torque, $\tau^*$, in Fig. \ref{fig.torque_axis_scale}(\emph{a}). We see that $\tau^*$ collapse reasonably well under the pitching axis scaling, despite some small discrepancies. The spanwise vorticity field, $\omega_z$, the $Q$ field, the influence field, $\phi$, and the moment density distribution, $-2Q\phi$, at $t/T=0$ are plotted in Fig. \ref{fig.torque_axis_scale}(\emph{b}) for further analysis of the pitching axis effect. The $\omega_z$ field shows that as the pitching axis moves from the mid-chord ($x/c=0.5$) to the quarter-chord ($x/c=0.25$), the leading-edge vortex becomes significantly weaker and less coherent, whereas the trailing-edge vortex becomes stronger and more coherent, due to a higher feeding shear-layer velocity (see Fig. \ref{fig.circ_axis_scale}). The $\phi$ field also changes significantly from $x/c=0.5$ to 0.25. The quadrant pattern disappears and $\phi$ becomes entirely negative on the upper surface of the wing and positive on the lower surface. This change in $\phi$ also alters the moment density distribution. We see that the weak LEV, despite its positive vorticity, now induces a negative torque, which is opposite to that induced by its counterpart at $x/c=0.5$. The TEV-induced torque becomes a lot higher because both $Q$ and $\phi$ increase. As the pitching axis further moves to the leading edge ($x/c=0$), a negative LEV is generated due to the strong pitch-induced flow around the leading edge. The influence field, $\phi$, stays similar to that of $x/c=0.25$ with an increase in magnitude. The negative LEV induces a positive torque, because it is on the upper surface of the wing. These complex behaviors of the pitch-induced vortices as the pitching axis shifts might account for the discrepancies observed in $\tau^*$ in Fig. \ref{fig.torque_axis_scale}(\emph{a}).

\begin{figure*}
\centering
\includegraphics[width=0.9\textwidth]{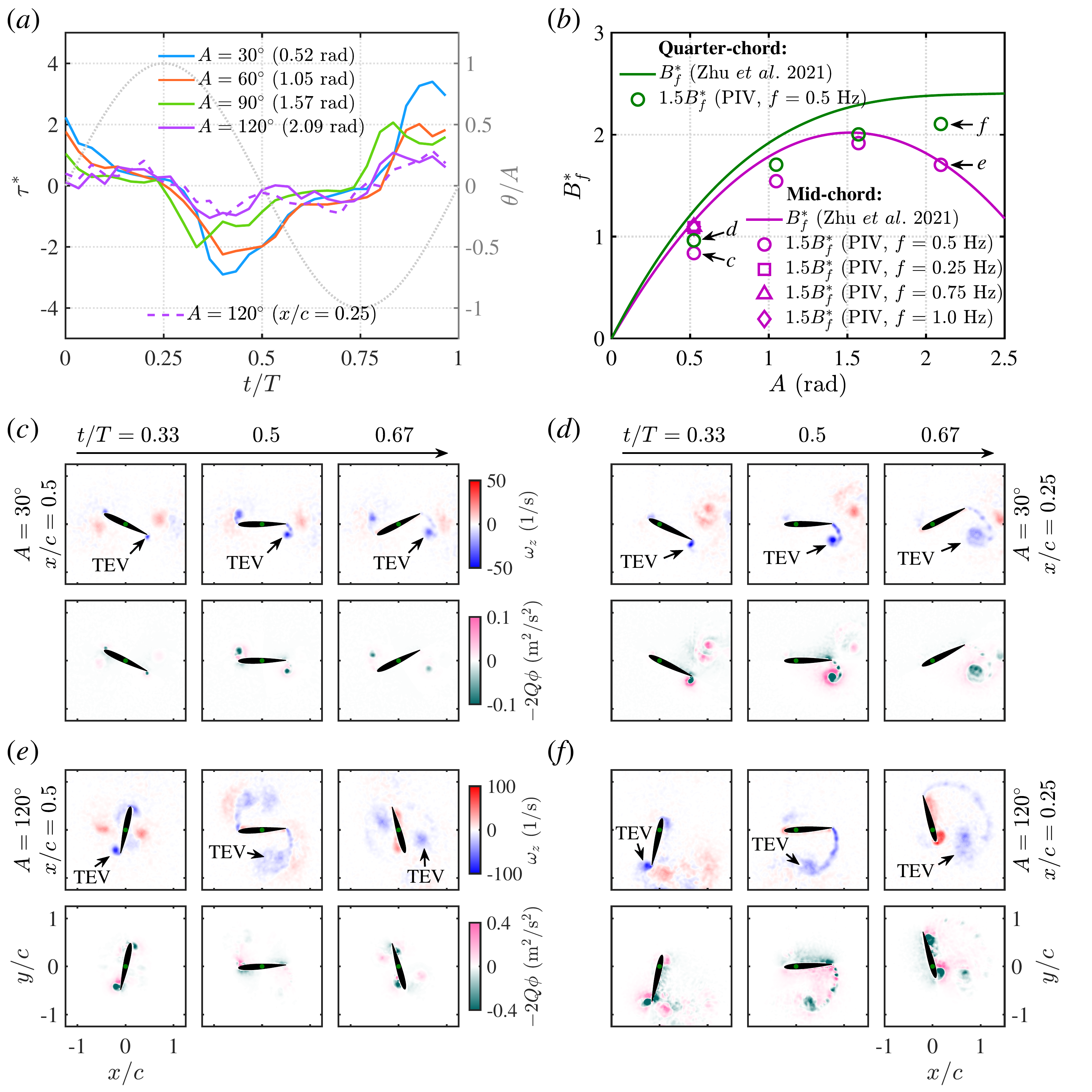}
\caption{(\emph{a}) Non-dimensional vorticity-induced torque ($\tau^*$) for a wing pitching at $x/c=0.5$, $f=0.5$ Hz and $A=30^\circ$ to $120^\circ$ (solid lines). Purple dashed line: $A=120^\circ$, $f=0.5$ Hz, $x/c=0.25$. (\emph{b}) Cycle-averaged non-dimensional fluid damping coefficient ($B^*_f$) extracted by ring-down experiments \citep[solid curves,][]{zhu2021nonlinear} and by PIV-based FMPM (hollow markers) for mid-chord pitching ($x/c=0.5$, purple) and quarter-chord pitching ($x/c=0.25$, green). Note the factor 1.5 applied to the FMPM data. The labeled data points correspond to (\emph{c--f}) temporal snapshots of the spanwise vorticity field ($\omega_z$, first row) and the moment density distribution ($-2Q\phi$, second row) during the pitch-down motion. (\emph{c}) $A=30^\circ$, $x/c=0.5$. (\emph{d}) $A=30^\circ$, $x/c=0.25$. (\emph{e}) $A=120^\circ$, $x/c=0.5$. (\emph{f}) $A=120^\circ$, $x/c=0.25$. The pitching frequency is maintained at $f=0.5$ Hz for all the cases.}
\label{fig.amp_scale_damp}
\end{figure*}

In Eqn. \ref{eqn.moment_scale}, we see that the vorticity-induced torque, $\tau$, scales with the pitching amplitude squared. However, this scaling is based on the linear assumption that trajectories of shed vortices stay perpendicular to the wing chord \citep{zhu2021nonlinear}, so it is presumably only valid for small-amplitude pitching. As the vortex trajectories vary nonlinearly for high pitching amplitudes (see Fig. \ref{fig.circ_amp_scale}\emph{a} inset), the scaling breaks down. This is confirmed by Fig. \ref{fig.amp_scale_damp}(\emph{a}), where we show that the non-dimensional vorticity-induced torque, $\tau^*$, does not collapse satisfactorily under the $A^2$ scaling, although the general trend of $\tau^*$ roughly matches for different pitching amplitudes. To further characterize the effect of pitching amplitudes on the vorticity-induced torque, in addition to the time trace data shown in Fig. \ref{fig.amp_scale_damp}(\emph{a}), we also look at the cycle-averaged $\tau^*$ and associate it with the non-dimensional fluid damping coefficient, $B_f^*=4Af\tau^*/\dot{\theta}$. In Fig. \ref{fig.amp_scale_damp}(\emph{b}), we compare $B_f^*$ obtained from the PIV-based FMPM (hollow markers) to those extracted by ``ring-down'' (direct torque) measurements of \citet{zhu2021nonlinear} (solid curves) as a function of the pitching amplitude. Two pitching axes are considered, as $B_f^*$ has been shown to behave differently when the wing pitches at the mid-chord ($x/c=0.5$) and the quarter-chord ($x/c=0.25$). The first thing we notice is that the PIV-based FMPM underestimates the vorticity-induced torque and hence the corresponding fluid damping. The potential cause for this underestimation will be discussed later. To better compare the trend between the FMPM-based $B_f^*$ and the ring-down-based $B_f^*$, we multiply a factor of 1.5 to the former. We see that the FMPM-based $B_f^*$ agrees very well in trend with those extracted by ring-down experiments. For $x/c=0.5$, $B_f^*$ increases non-monotonically with the pitching amplitude, whereas for $x/c=0.25$, $B_f^*$ increases monotonically with the pitching amplitude with a decreasing slope.

To explain the differences in $B_f^*$ for different pitching amplitudes and axes, we choose four representative cases (data points \emph{c--f} on Fig. \ref{fig.amp_scale_damp}\emph{b}) and plot the corresponding spanwise vorticity field, $\omega_z$, and the moment density distribution, $-2Q\phi$, in Fig. \ref{fig.amp_scale_damp}(\emph{c--f}). For each case, three temporal snapshots, $t/T=0.33$, 0.5 and 0.67, are plotted to capture the initial, middle and late stages of the pitch-down motion. Fig. \ref{fig.amp_scale_damp}(\emph{c}) depicts a conventional scenario where the wing pitches at $A=30^\circ$ and $x/c=0.5$. In this case, two negative vortices are generated at the leading edge and the trailing edge. These two vortices are of comparable size and strength (see also Fig. \ref{fig.torque_LETE_f_scale}\emph{a}), and both contribute to negative moments. When the pitching axis moves to $x/c=0.25$ (Fig. \ref{fig.amp_scale_damp}\emph{d}), the LEV becomes weaker and the TEV becomes stronger, due to the change in the feeding shear-layer velocities. While the LEV is still negative, it generates a small positive moment due to the negative $\phi$ field (see Fig. \ref{fig.torque_axis_scale}\emph{b}).

Comparing Fig. \ref{fig.amp_scale_damp}(\emph{e}) to (\emph{c}), we see that when the pitching amplitude is very high ($A=120^\circ$), the LEV and TEV both move towards the pitching axis from $t/T=0.33$ to 0.5. At the same time, they also become less coherent. These two effects combined lead to the near-zero $\tau^*$ observed in Fig. \ref{fig.amp_scale_damp}(\emph{a}) at $t/T=0.5$. As the pitch-down motion continues, the LEV moves to the aft wing, and the TEV moves to the fore wing, resulting in a sign switch of the induced torque -- the LEV and TEV both generate positive moments at the late stage of the motion ($t/T=0.67$). However, the total vorticity-induced torque, $\tau^*$ remains slightly negative because of the positive surface vortices, which are closer to the wing surface and thus generate more negative moments. The fact that the LEV and TEV move across the pitching axis brings $\tau^*$ close to zero earlier than that of smaller pitching amplitudes. This further results in the decreasing $B_f^*$ observed for mid-chord pitching wings at large pitching amplitudes. At this same pitching amplitude ($A=120^\circ$), as the pitching axis moves to $x/c=0.25$ (Fig. \ref{fig.amp_scale_damp}\emph{f}), we see that the TEV again moves across the pitching axis at the late stage ($t/T=0.67$). However, because the $\phi$ field is entirely positive under the wing (see Fig. \ref{fig.torque_axis_scale}\emph{b}), the TEV continues to generate negative moments. In addition, a positive LEV emerges due to the pitch-induced flow and also generates a negative moment. These effects result in a higher $\tau^*$ magnitude at $t/T>0.5$ for $x/c=0.25$ (Fig. \ref{fig.amp_scale_damp}\emph{a}, purple dashed line) as compared to the $x/c=0.5$ case (purple solid line), explaining the difference we see in $B_f^*$ at $A=120^\circ$ (Fig. \ref{fig.amp_scale_damp}\emph{b}, data points \emph{f} and \emph{e}). We want to note that, although the LEVs and TEVs move across the pitching axis for large-amplitude pitching, the vorticity-induced moment always stays negative during the pitch-down motion ($0.25<t/T<0.75$), where the angular velocity is also negative ($\dot{\theta}<0$). This assures that the instantaneous aerodynamic damping is always positive (i.e. $b_f=\tau/\dot{\theta}>0$) during the entire pitching cycle, which holds valid for all the cases considered in the present study (see Fig. \ref{fig.torque_LETE_f_scale}\emph{a}, Fig. \ref{fig.torque_axis_scale}\emph{a} and Fig. \ref{fig.amp_scale_damp}\emph{a}).

\begin{figure*}
\centering
\includegraphics[width=0.9\textwidth]{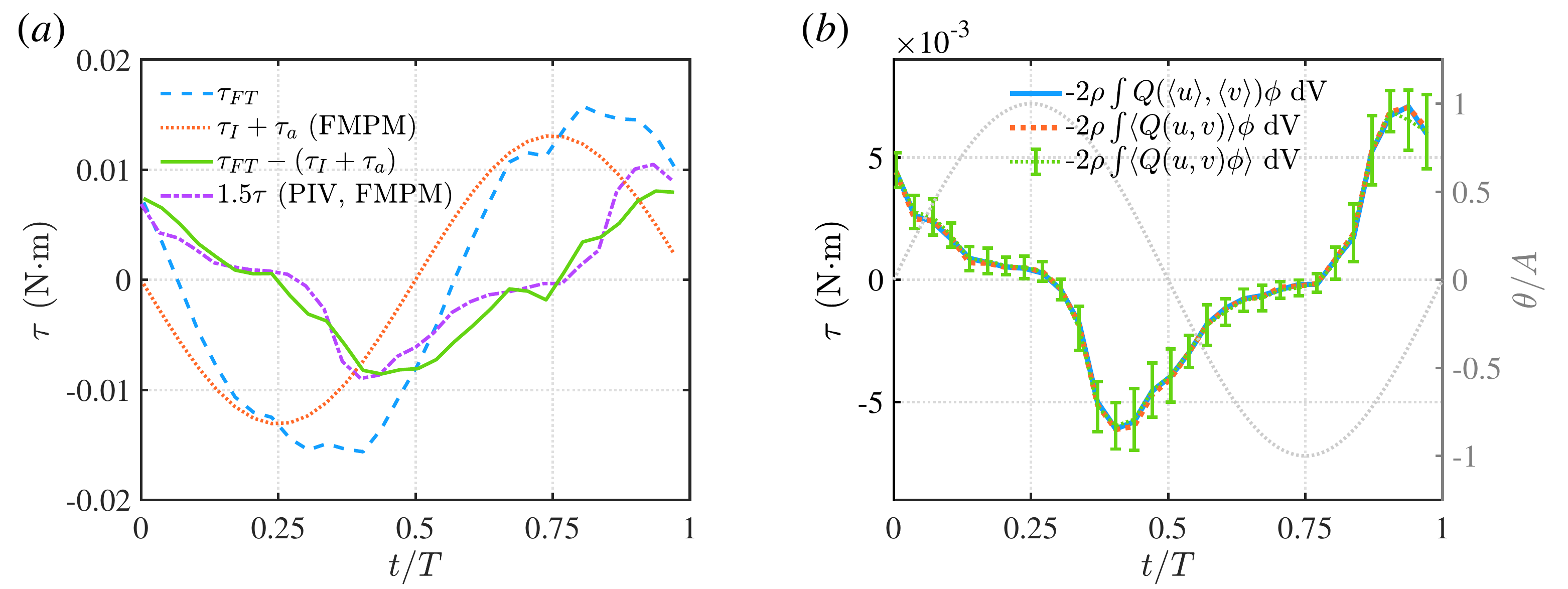}
\caption{(\emph{a}) Time trace of the torque measured by the force transducer ($\tau_{FT}$), the sum of the inertial torque and the added-mass torque calculated using FMPM ($\tau_I+\tau_a$), the true vorticity-induced torque ($\tau_{FT}-(\tau_I+\tau_a)$) and 1.5 times the vorticity-induced torque calculated using PIV-based FMPM (1.5$\tau$) for $A=30^\circ$, $f=0.5$ Hz, $x/c=0.5$. (\emph{b}) vorticity-induced torque, $\tau$, based on $Q$ calculated using phase-averaged velocity fields ($\langle{u}\rangle,\langle{v}\rangle$), phase-averaged $\langle{Q}\rangle$ calculated using instantaneous velocity fields ($u,v$), and phase-averaged $\langle{Q\phi}\rangle$ calculated using instantaneous velocity fields ($u,v$). Error bars denote standard deviations over 20 cycles.}
\label{fig.phase-avg_compare}
\end{figure*}

\subsection{Error analysis of FMPM results}{\label{sec.error}}

To further explain the underestimation of the fluid damping coefficient by the FMPM (Fig. \ref{fig.amp_scale_damp}\emph{b}), we compare the vorticity-induced torque calculated by the PIV-based FMPM to that measured by the force transducer (see Fig. \ref{fig.setup}\emph{a}). According to the Morison equation \citep{morison1950force}, the total fluid force experienced by a moving body can be divided into the vorticity-induced force and the added-mass force. Therefore, to get the vorticity-induced torque, $\tau$, from the force transducer, we have to first estimate the added-mass torque, $\tau_a$, using Eqn. \ref{eqn.added_mass}. Then, this torque as well as the physical wing inertia, $\tau_I$, are subtracted from the measured torque, $\tau_{FT}$. The inertial torque, $\tau_I$, was measured from experiments conducted in air.

The torque measured by the force transducer, $\tau_{FT}$, the sum of the inertial torque and the added-mass torque, $\tau_I+\tau_a$, the true vorticity-induced torque, $\tau_{FT}-(\tau_I+\tau_a)$, and the vorticity-induced torque calculated using PIV-based FMPM, $\tau$, for $A=30^\circ$, $f=0.5$ Hz, $x/c=0.25$ are plotted in Fig. \ref{fig.phase-avg_compare}(\emph{a}). We assume that the viscous torque is negligible in comparison to the vortex-induced torque, due to the relatively high Reynolds number - $Re\sim\mathcal{O}(10^4)$. We find that the PIV-based FMPM underestimates the vorticity-induced torque, $\tau$, roughly by a factor of 1.5, which explains the 1.5 factor used for the non-dimensional fluid damping coefficient, $B_f^*$, in Fig. \ref{fig.amp_scale_damp}(\emph{b}). After applying a factor of 1.5 to $\tau$, we see that it agrees well with the true vorticity-induced torque, $\tau_{FT}-(\tau_I+\tau_a)$.

However, the question remains why the PIV-based FMPM significantly underestimates the vorticity-induced torque. One conjecture, as discussed earlier, is that because we are using phase-averaged PIV velocity fields ($\langle{u}\rangle,\langle{v}\rangle$) to calculate the $Q$ fields, some small instantaneous flow structures, which also contribute to the moment generation, might have been averaged out, as discussed in Section \ref{sec.phase_avg}. To assess this effect, we compare $\tau$ based on $Q$ calculated using phase-averaged velocity fields ($\langle{u}\rangle,\langle{v}\rangle$), phase-averaged $\langle{Q}\rangle$ calculated using instantaneous velocity fields ($u,v$), and phase-averaged $\langle{Q\phi}\rangle$ calculated using instantaneous velocity fields ($u,v$) in Fig. \ref{fig.phase-avg_compare}(\emph{b}). We see that the vorticity-induced torque calculated using these three different methods matches closely, indicating that phase-averaging is \textit{not} the main cause for the FMPM to underestimate $\tau$. The good agreements between $-2\rho \int Q(\langle{u}\rangle,\langle{v}\rangle)\phi \mathrm{dV}$ (blue solid curve) and $-2\rho \int \langle{Q}(u,v)\rangle\phi \mathrm{dV}$ (orange dashed curve) also indicates that the $2Q_{\bar{u}\langle{u}\rangle}$ term in Eqn. \ref{eqn.phase_avg} might be dropped for the cases considered in the present study.

Another potential error source comes from the PIV measurements, and in particular, the difficulty in obtaining accurate velocity vectors near the solid boundary \citep{rival2017load}. Because the vorticity-induced torque is calculated by integrating the $-2Q\phi$ field (Eqn. \ref{eqn.vortex_torque}), any missing velocity vectors near the solid boundary will result in a significant decrease of the overall vorticity-induced torque, as $\phi$ reaches its maximum near the boundary. This conjecture could be tested by comparing the PIV-measured near-boundary velocity fields with those obtained by from a high-accuracy numerical simulation. Alternatively, a physics-informed neural network, PINN \citep{raissi2019physics,cai2021flow,arzani2021uncovering}, could potentially be used to reconstruct and resolve near-boundary velocity fields so as to improve the accuracy of the PIV-based FMPM. These tasks, unfortunately, lie beyond the scope of the current paper but are well worthy of investigation.

Lastly, all FMPM calculations in this study are based on two-component, two-dimensional (2C2D) PIV measurements taken at the mid span, but without considering three-dimensional effects. This means the contributions from the spanwise velocity ($w$) and the spanwise gradient ($\partial/\partial z$) of ($u,v$) to $Q$ have not been considered, but might play an important role in causing the differences between the PIV-based FMPM results and the force sensor results. The recent paper of \citet{menon2022contribution} has employed FMPM to quantify the role of cross-span vorticity on the force generation over a finite-aspect ratio wing and these effects were shown to be quite significant. In addition, the pitching wing in this experiment has a free wingtip (Fig. \ref{fig.setup}\emph{a}) and the tip vortex presumably plays a non-negligible role in the generation of the total torque. As with the boundary-related errors, one could assess the three-dimensional flow effects by comparing our results with 2D and 3D CFD simulations or by conducting 3D PIV measurements, both of which, unfortunately, are beyond the scope of the present study.

\section{Conclusion}\label{sec.conclusion}

In this study, we have characterized the vortex dynamics associated with a NACA 0012 wing undergoing prescribed sinusoidal pitching in quiescent water. We employed two-dimensional particle image velocimetry (PIV) to measure the velocity field around the wing, and used the $Q$ criterion to identify positions and boundaries of pitch-generated vortices to study the evolution of their trajectories and strengths. We found that the vortex trajectory is insensitive to the pitching frequency, but varies nonlinearly with the pitching amplitude and scales linearly with the pitching axis. The vortex circulation was shown to scale with the pitching frequency, amplitude, and effective chord length squared for sinusoidal pitch oscillations. A vortex splitting behavior causing the vortex circulation to drop abruptly after a linear growth regime was observed for all the pitching cases, except for those at the mid-chord and the lowest pitching amplitude.

In the second part of this study, the Force and Moment Partitioning Method (FMPM) was adopted to quantify and visualize the aerodynamic moment generated by the pitch-induced vortices. The moment contributions from leading-edge vortices and trailing-edge vortices were separated by the FMPM, and the ratio between the two was shown to match the empirical factor used in \citet{zhu2021nonlinear}. A scaling for the vorticity-induced torque was proposed, revealing its dependence on the squared pitching frequency, the squared pitching amplitude, and the fourth power of the effective chord length. The pitching amplitude was shown to have a nonlinear effect on the vorticity-induced moment due to the complex vortex dynamics. The vorticity-induced moment was further connected with the fluid damping reported by \citet{zhu2021nonlinear}, and the results obtained using PIV-based FMPM were found to match well with that measured using ring-down experiments, despite a lower magnitude. Finally, the FMPM was found to underestimate the moment compared to the force transducer data, potentially due to the missing velocity vectors near the wing boundary and three-dimensional effects. This error is a concern and will be addressed in follow-up studies.

Together with our previous study on cycle-averaged vorticity-induced damping \citep{zhu2021nonlinear}, the present work, which focuses on the instantaneous evolution of vortex dynamics and moments, provides a comprehensive understanding of the frequency, amplitude and pivot axis effects on the trajectory, strength, and associated aerodynamic moment of vortices shed from a sinusoidally pitching wing in quiescent water, a configuration that is of tremendous engineering and biological relevance. Moreover, this work is among the first to apply FMPM for analyzing experimental data \citep[see also][]{kumar2021data}. The good agreements we see between the FMPM-based results and the ring-down experiments/force transducer measurements further demonstrate the effectiveness and robustness of this method. The discussions on applying FMPM to phase-averaged data and the possible error source for causing the underestimation of the vorticity-induced moment can potentially benefit future applications of FMPM to experimental data.

\section*{Declarations}

\textbf{Ethical Approval}\\
Not applicable.

\vspace{0.5em}
\noindent\textbf{Competing interests}\\
The authors have no conflicts of interest to declare that are relevant to the content of this article.

\vspace{0.5em}
\noindent\textbf{Authors' contributions}\\
All authors contributed to the work. YZ and KB conceived and designed the experiments. YZ conducted the experiments and collected the data. YZ, HL and KB analyzed the force and PIV data. YZ, SK, KM, RM and KB performed the FMPM analysis. YZ and KB wrote the manuscript. All authors reviewed and improved the manuscript.

\vspace{0.5em}
\noindent\textbf{Funding}\\
This work was funded by the Air Force Office of Scientific Research, Grant FA9550-21-1-0462, managed by Dr. Gregg Abate. RM acknowledges support from NSF grant CBET-2011619 and ONR Grants N00014-22-1-2655 and N00014-22-1-2770 monitored by Dr. Bob Brizzolara.

\vspace{0.5em}
\noindent\textbf{Availability of data and materials}\\
The datasets generated during and/or analyzed during the current study are available from the corresponding author upon reasonable request.

\begin{appendices}

\section{Universal scaling of the vortex circulation}

\begin{figure}[ht]
\centering
\includegraphics[width=0.475\textwidth]{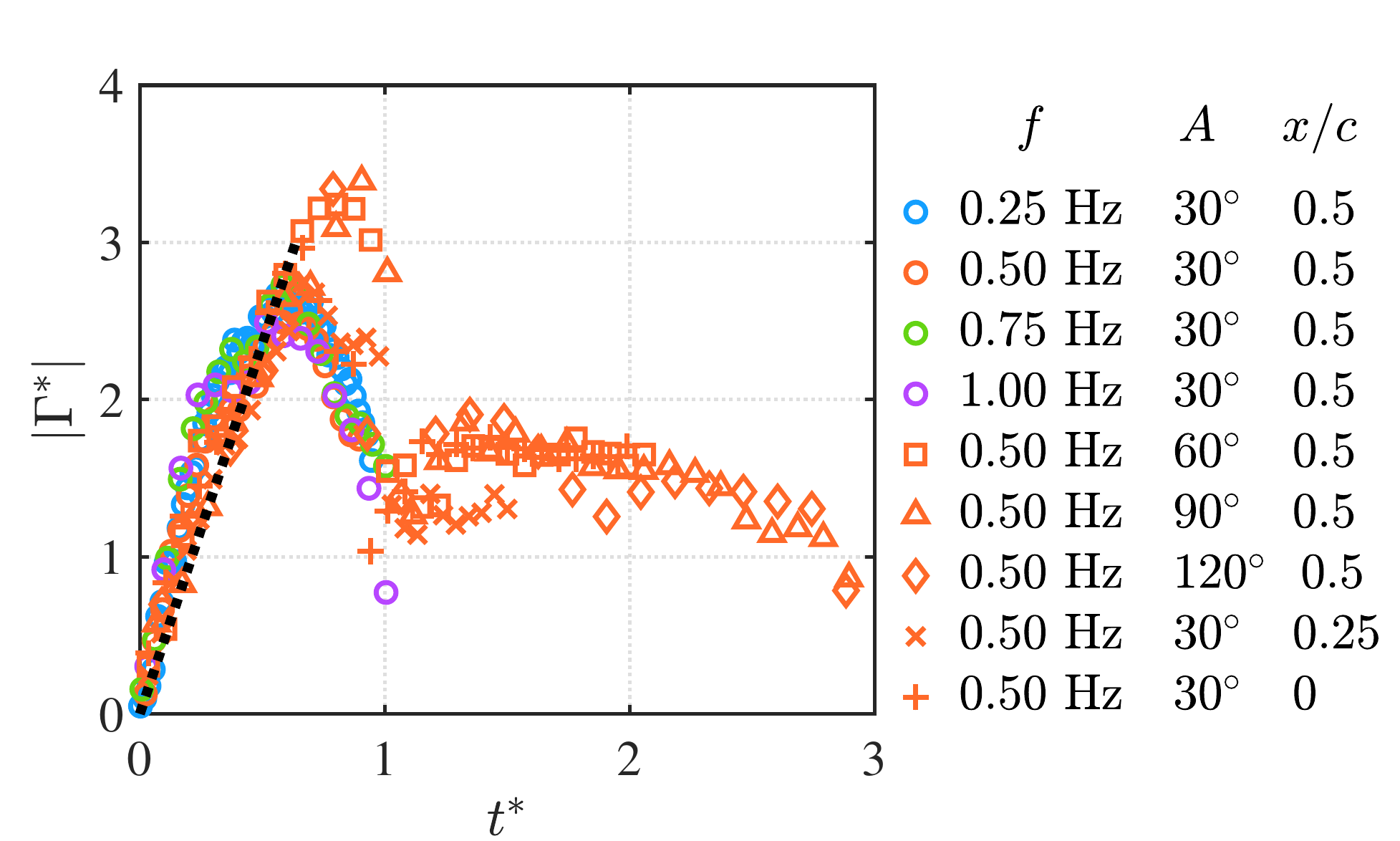}
\caption{Non-dimensional vortex circulation as a function of the non-dimensional vortex formation time across a range of pitching frequency, amplitude and pivot axis. Data are replotted from Figs. \ref{fig.circ_f_scale}-\ref{fig.circ_axis_scale}(\emph{b}).}
\label{fig.universal}
\end{figure}

In Section \ref{sec.traj_circ}, we discussed the effect of pitching frequency, amplitude, and pivot axis on the vortex strength and trajectory. To better demonstrate the effect of each individual parameter, we plot $\Gamma^*$ as a function of $t^*$ in three figures (Figs. \ref{fig.circ_f_scale}-\ref{fig.circ_axis_scale}\emph{b}) separately. However, it is also important to note that the $\Gamma^*\sim t^*$ scaling is universal (see also \citet{onoue2016vortex}). To demonstrate this universality, we replot all the data from Figs. \ref{fig.circ_f_scale}-\ref{fig.circ_axis_scale}(\emph{b}) in Fig. \ref{fig.universal}. We see that the data from different frequencies, amplitudes and pivot axes collapse well under the $\Gamma^*\sim t^*$ scaling, despite the slight discrepancy in slope between the small- and large-amplitude cases.

\section{Other considerations on the calculation of the vorticity-induced moment}
\subsection{Effect of the integration domain}

\begin{figure}[ht]
\centering
\includegraphics[width=0.45\textwidth]{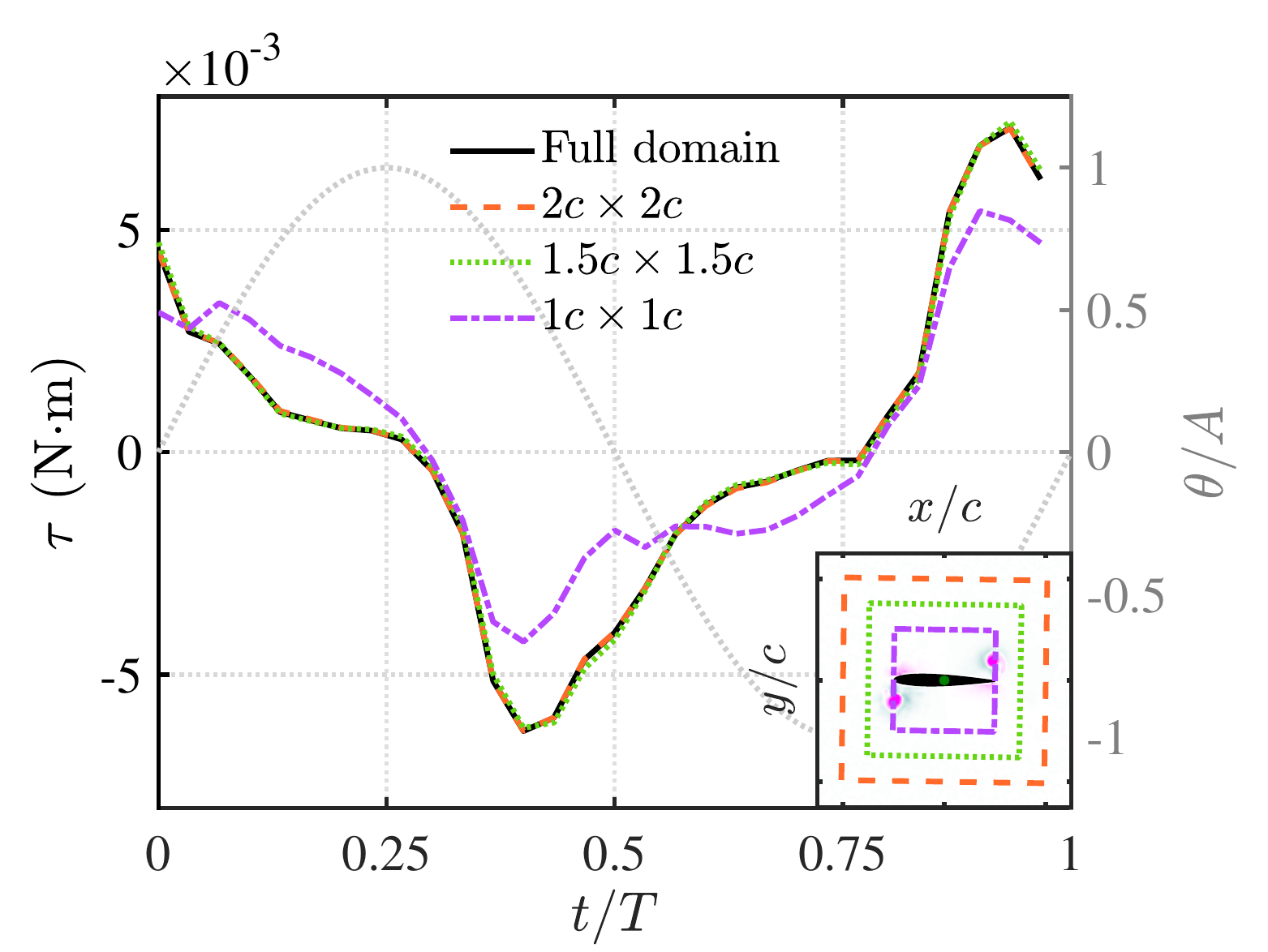}
\caption{Effect of integration domain on the calculation of vorticity-induced torque ($A=30^\circ$, $f=0.5$ Hz, $x/c=0.5$).}
\label{fig.domain}
\end{figure}

Vorticities leaving the PIV (or the integration) domain may cause an underestimation of the vorticity-induced force/moment. However, this effect may not be significant because the influence potential decreases rapidly moving away from the wing surface. To further characterize this domain effect, we manually shrink the integration domain and recalculate the vorticity-induced moment. We consider a low-amplitude pitching case ($A=30^\circ$, $f=0.5$ Hz, $x/c=0.5$), and observe that reducing the integration window has a minimal effect on the vorticity-induced moment, except for the smallest integration window ($1c \times 1c$). This is partially because the shed vortices stay near the wing for small-amplitude pitching. Reducing the integration domain may have a stronger effect on cases with higher pitching amplitudes. However, we believe this effect should not be a concern for the present study, where a relatively large PIV domain has been used.

\subsection{Effect of the vector field resolution}

\begin{figure}[ht]
\centering
\includegraphics[width=0.45\textwidth]{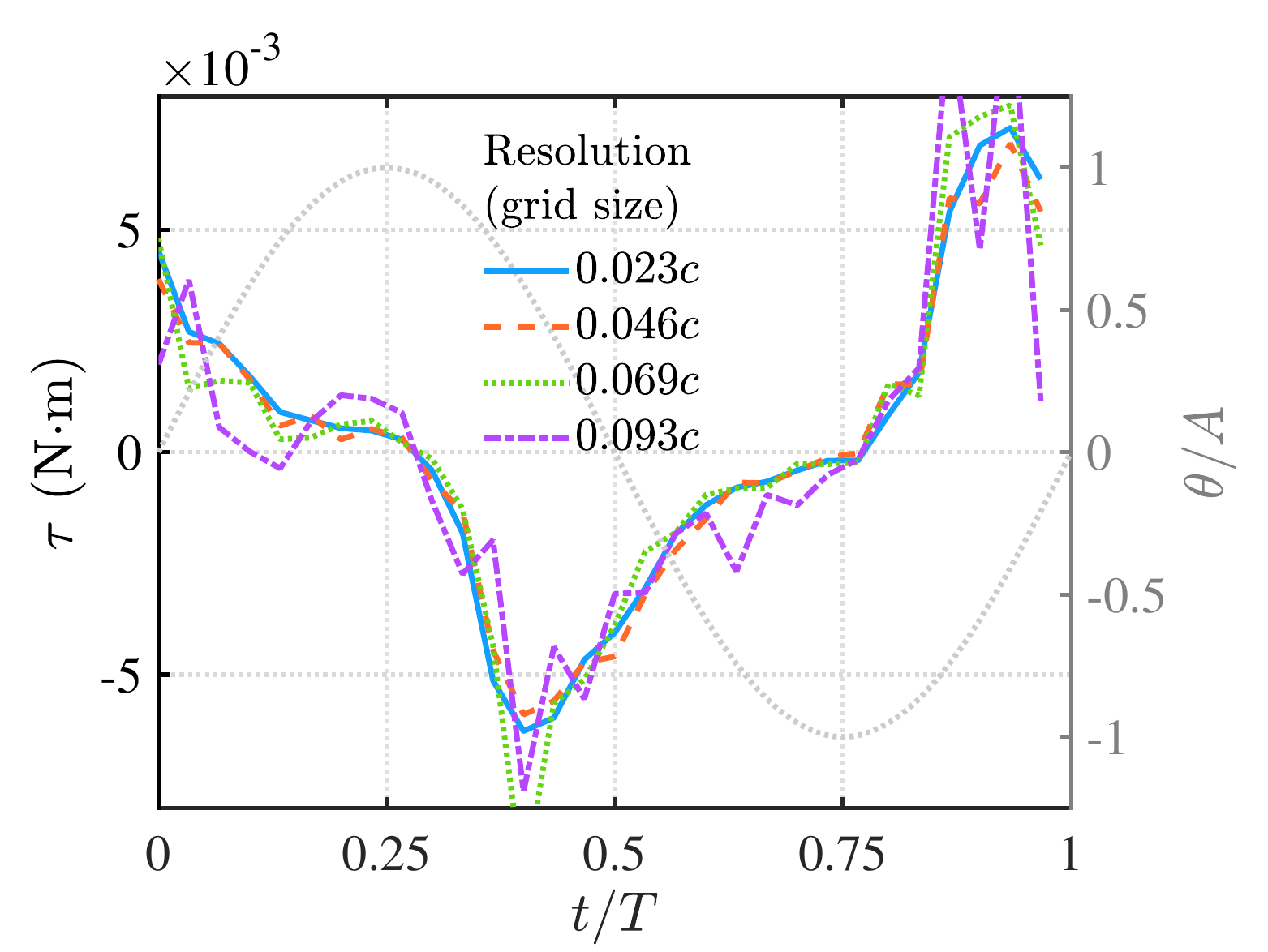}
\caption{Effect of data resolution on the calculation of vorticity-induced torque ($A=30^\circ$, $f=0.5$ Hz, $x/c=0.5$).}
\label{fig.spacing}
\end{figure}

Data spacing may have an effect on the vorticity-induced moment calculated using FMPM. To evaluate this effect, we manually reduce the resolution of the PIV data from $0.023c$ to $0.093c$ and recalculate the vorticity-induced moment for the case of $A=30^\circ$, $f=0.5$ Hz, $x/c=0.5$. We see that the vorticity-induced moment stays almost unchanged when the resolution is reduced to half ($0.046c$). Variations start to emerge when the resolution is further reduced ($0.069c$), and the vorticity-induced moment becomes noisy at the lowest resolution ($0.093c$). This resolution test shows that higher spatial resolution is desired for FMPM calculations, and more than $\sim 20$ vectors per chord (higher than $\sim 0.05c$ resolution) is preferred.

\end{appendices}

\bibliography{reference}

\begin{thebibliography}{61}
\providecommand{\natexlab}[1]{#1}
\providecommand{\url}[1]{{#1}}
\providecommand{\urlprefix}{URL }
\providecommand{\doi}[1]{\url{https://doi.org/#1}}
\providecommand{\eprint}[2][]{\url{#2}}
 \bibcommenthead

\bibitem[{Anderson et~al(1998)Anderson, Streitlien, Barrett, and
  Triantafyllou}]{anderson1998oscillating}
Anderson JM, Streitlien K, Barrett DS, et~al (1998) Oscillating foils of high
  propulsive efficiency. J Fluid Mech 360:41--72

\bibitem[{Arzani et~al(2021)Arzani, Wang, and D'Souza}]{arzani2021uncovering}
Arzani A, Wang JX, D'Souza RM (2021) Uncovering near-wall blood flow from
  sparse data with physics-informed neural networks. Phys Fluids 33(7):071,905

\bibitem[{Baik et~al(2012)Baik, Bernal, Granlund, and Ol}]{baik2012unsteady}
Baik YS, Bernal LP, Granlund K, et~al (2012) Unsteady force generation and
  vortex dynamics of pitching and plunging aerofoils. J Fluid Mech 709:37--68

\bibitem[{Barrett(1996)}]{barrett1996propulsive}
Barrett DS (1996) Propulsive efficiency of a flexible hull underwater vehicle.
  PhD thesis, Massachusetts Institute of Technology

\bibitem[{Bergou et~al(2007)Bergou, Xu, and Wang}]{bergou2007passive}
Bergou AJ, Xu S, Wang ZJ (2007) Passive wing pitch reversal in insect flight. J
  Fluid Mech 591:321--337

\bibitem[{Buchholz and Smits(2008)}]{buchholz2008wake}
Buchholz JHJ, Smits AJ (2008) The wake structure and thrust performance of a
  rigid low-aspect-ratio pitching panel. J Fluid Mech 603:331--365

\bibitem[{Buchholz et~al(2011)Buchholz, Green, and Smits}]{buchholz2011scaling}
Buchholz JHJ, Green MA, Smits AJ (2011) Scaling the circulation shed by a
  pitching panel. J Fluid Mech 688:591--601

\bibitem[{Cai et~al(2021)Cai, Wang, Fuest, Jeon, Gray, and
  Karniadakis}]{cai2021flow}
Cai S, Wang Z, Fuest F, et~al (2021) Flow over an espresso cup: inferring 3-d
  velocity and pressure fields from tomographic background oriented schlieren
  via physics-informed neural networks. J Fluid Mech 915:A102

\bibitem[{Corke and Thomas(2015)}]{corke2015dynamic}
Corke TC, Thomas FO (2015) Dynamic stall in pitching airfoils: aerodynamic
  damping and compressibility effects. Annu Rev Fluid Mech 47:479--505

\bibitem[{Corkery et~al(2019)Corkery, Babinsky, and
  Graham}]{corkery2019quantification}
Corkery SJ, Babinsky H, Graham WR (2019) Quantification of added-mass effects
  using particle image velocimetry data for a translating and rotating flat
  plate. J Fluid Mech 870:492--518

\bibitem[{Dabiri(2009)}]{dabiri2009optimal}
Dabiri JO (2009) Optimal vortex formation as a unifying principle in biological
  propulsion. Annu Rev Fluid Mech 41:17--33

\bibitem[{DeVoria and Ringuette(2012)}]{devoria2012vortex}
DeVoria AC, Ringuette MJ (2012) Vortex formation and saturation for
  low-aspect-ratio rotating flat-plate fins. Exp Fluids 52(2):441--462

\bibitem[{DeVoria and Ringuette(2013)}]{devoria2013force}
DeVoria AC, Ringuette MJ (2013) The force and impulse of a flapping plate
  performing advancing and returning strokes in a quiescent fluid. Exp Fluids
  54(5):1--15

\bibitem[{Ellington et~al(1996)Ellington, van~den Berg, Willmott, and
  Thomas}]{ellington1996leading}
Ellington CP, van~den Berg C, Willmott AP, et~al (1996) Leading-edge vortices
  in insect flight. Nature 384(6610):626

\bibitem[{Epps and Techet(2007)}]{epps2007impulse}
Epps BP, Techet AH (2007) Impulse generated during unsteady maneuvering of
  swimming fish. Exp Fluids 43(5):691--700

\bibitem[{Francescangeli and Mulleners(2021)}]{francescangeli2021discrete}
Francescangeli D, Mulleners K (2021) Discrete shedding of secondary vortices
  along a modified kaden spiral. J Fluid Mech 917:A44

\bibitem[{Gehlert and Babinsky(2021)}]{gehlert2021noncirculatory}
Gehlert P, Babinsky H (2021) Noncirculatory force on a finite thickness body
  encountering a gust. AIAA J 59(2):719--730

\bibitem[{Gharib et~al(1998)Gharib, Rambod, and Shariff}]{gharib1998universal}
Gharib M, Rambod E, Shariff K (1998) A universal time scale for vortex ring
  formation. J Fluid Mech 360:121--140

\bibitem[{Heathcote et~al(2004)Heathcote, Martin, and
  Gursul}]{heathcote2004flexible}
Heathcote S, Martin D, Gursul I (2004) Flexible flapping airfoil propulsion at
  zero freestream velocity. AIAA J 42(11):2196--2204

\bibitem[{Ho et~al(2003)Ho, Nassef, Pornsinsirirak, Tai, and
  Ho}]{ho2003unsteady}
Ho S, Nassef H, Pornsinsirirak N, et~al (2003) Unsteady aerodynamics and flow
  control for flapping wing flyers. Prog Aerosp Sci 39(8):635--681

\bibitem[{Hunt et~al(1988)Hunt, Wray, and Moin}]{hunt1988eddies}
Hunt JCR, Wray AA, Moin P (1988) Eddies, streams, and convergence zones in
  turbulent flows. Center for Turbulence Research Report CTR-S88 pp 193--208

\bibitem[{Jafferis et~al(2019)Jafferis, Helbling, Karpelson, and
  Wood}]{jafferis2019untethered}
Jafferis NT, Helbling EF, Karpelson M, et~al (2019) Untethered flight of an
  insect-sized flapping-wing microscale aerial vehicle. Nature
  570(7762):491--495

\bibitem[{Jeong and Hussain(1995)}]{jeong1995identification}
Jeong J, Hussain F (1995) On the identification of a vortex. J Fluid Mech
  285:69--94

\bibitem[{Jimreeves~David et~al(2018)Jimreeves~David, Mathur, Govardhan, and
  Arakeri}]{david2018kinematic}
Jimreeves~David M, Mathur M, Govardhan RN, et~al (2018) The kinematic genesis
  of vortex formation due to finite rotation of a plate in still fluid. J Fluid
  Mech 839:489--524

\bibitem[{Kang and Shyy(2014)}]{kang2014analytical}
Kang CK, Shyy W (2014) Analytical model for instantaneous lift and shape
  deformation of an insect-scale flapping wing in hover. J R Soc Interface
  11(101):20140,933

\bibitem[{Kumar et~al(2021)Kumar, Brooks, Green, and Mittal}]{kumar2021data}
Kumar S, Brooks S, Green M, et~al (2021) A data-driven method for determining
  the hydrodynamic force induced by vortices-force partitioning applied to piv
  data for a caudal fin model. In: APS DFD Meeting Abstracts, pp M13--003

\bibitem[{Lee et~al(2022)Lee, Simone, Su, Zhu, Ribeiro, Franck, and
  Breuer}]{lee2022leading}
Lee H, Simone N, Su Y, et~al (2022) Leading edge vortex formation and wake
  trajectory: Synthesizing measurements, analysis, and machine learning. Phys
  Rev Fluids 7:074,704

\bibitem[{Lentink and Dickinson(2009)}]{lentink2009rotational}
Lentink D, Dickinson MH (2009) Rotational accelerations stabilize leading edge
  vortices on revolving fly wings. J Exp Biol 212(16):2705--2719

\bibitem[{Li and Wu(2018)}]{li2018vortex}
Li J, Wu ZN (2018) Vortex force map method for viscous flows of general
  airfoils. J Fluid Mech 836:145--166

\bibitem[{Li et~al(2020)Li, Wang, Graham, and Zhao}]{li2020vortex}
Li J, Wang Y, Graham M, et~al (2020) Vortex moment map for unsteady
  incompressible viscous flows. J Fluid Mech 891:A13

\bibitem[{McCroskey(1982)}]{mccroskey1982unsteady}
McCroskey WJ (1982) Unsteady airfoils. Annu Rev Fluid Mech 14(1):285--311

\bibitem[{Menon and Mittal(2019)}]{menon2019flow}
Menon K, Mittal R (2019) Flow physics and dynamics of flow-induced pitch
  oscillations of an airfoil. J Fluid Mech 877:582--613

\bibitem[{Menon and Mittal(2020)}]{menon2020dynamic}
Menon K, Mittal R (2020) Dynamic mode decomposition based analysis of flow over
  a sinusoidally pitching airfoil. J Fluids Struct 94:102,886

\bibitem[{Menon and Mittal(2021{\natexlab{a}})}]{menon2021initiation}
Menon K, Mittal R (2021{\natexlab{a}}) On the initiation and sustenance of
  flow-induced vibration of cylinders: insights from force partitioning. J
  Fluid Mech 907

\bibitem[{Menon and Mittal(2021{\natexlab{b}})}]{menon2021quantitative}
Menon K, Mittal R (2021{\natexlab{b}}) Quantitative analysis of the kinematics
  and induced aerodynamic loading of individual vortices in vortex-dominated
  flows: a computation and data-driven approach. J Comput Phys 443:110,515

\bibitem[{Menon and Mittal(2021{\natexlab{c}})}]{menon2021significance}
Menon K, Mittal R (2021{\natexlab{c}}) Significance of the strain-dominated
  region around a vortex on induced aerodynamic loads. J Fluid Mech 918:R3

\bibitem[{Menon et~al(2022)Menon, Kumar, and Mittal}]{menon2022contribution}
Menon K, Kumar S, Mittal R (2022) Contribution of spanwise and cross-span
  vortices to the lift generation of low-aspect-ratio wings: Insights from
  force partitioning. Phys Rev Fluids 7(11):114,102

\bibitem[{Milano and Gharib(2005)}]{milano2005uncovering}
Milano M, Gharib M (2005) Uncovering the physics of flapping flat plates with
  artificial evolution. J Fluid Mech 534:403--409

\bibitem[{Moriche et~al(2017)Moriche, Flores, and
  Garc{\'\i}a-Villalba}]{moriche2017aerodynamic}
Moriche M, Flores O, Garc{\'\i}a-Villalba M (2017) On the aerodynamic forces on
  heaving and pitching airfoils at low reynolds number. J Fluid Mech
  828:395--423

\bibitem[{Morison et~al(1950)Morison, Johnson, and Schaaf}]{morison1950force}
Morison JR, Johnson JW, Schaaf SA (1950) The force exerted by surface waves on
  piles. J Pet Technol 2(05):149--154

\bibitem[{Morse and Williamson(2009)}]{morse2009prediction}
Morse TL, Williamson CHK (2009) Prediction of vortex-induced vibration response
  by employing controlled motion. J Fluid Mech 634:5--39

\bibitem[{Onoue and Breuer(2016)}]{onoue2016vortex}
Onoue K, Breuer K (2016) Vortex formation and shedding from a cyber-physical
  pitching plate. J Fluid Mech 793:229--247

\bibitem[{Onoue and Breuer(2017)}]{onoue2017scaling}
Onoue K, Breuer K (2017) A scaling for vortex formation on swept and unswept
  pitching wings. J Fluid Mech 832:697--720

\bibitem[{Quartapelle and Napolitano(1983)}]{quartapelle1983force}
Quartapelle L, Napolitano M (1983) Force and moment in incompressible flows.
  AIAA J 21(6):911--913

\bibitem[{Raissi et~al(2019)Raissi, Perdikaris, and
  Karniadakis}]{raissi2019physics}
Raissi M, Perdikaris P, Karniadakis GE (2019) Physics-informed neural networks:
  A deep learning framework for solving forward and inverse problems involving
  nonlinear partial differential equations. J Comput Phys 378:686--707

\bibitem[{Ringuette et~al(2007)Ringuette, Milano, and
  Gharib}]{ringuette2007role}
Ringuette MJ, Milano M, Gharib M (2007) Role of the tip vortex in the force
  generation of low-aspect-ratio normal flat plates. J Fluid Mech 581:453--468

\bibitem[{Rival et~al(2009)Rival, Prangemeier, and Tropea}]{rival2009influence}
Rival D, Prangemeier T, Tropea C (2009) The influence of airfoil kinematics on
  the formation of leading-edge vortices in bio-inspired flight. Exp Fluids
  46:823--833

\bibitem[{Rival and van Oudheusden(2017)}]{rival2017load}
Rival DE, van Oudheusden B (2017) Load-estimation techniques for unsteady
  incompressible flows. Exp Fluids 58(3):1--11

\bibitem[{Seo and Mittal(2022)}]{seo2022improved}
Seo JH, Mittal R (2022) Improved swimming performance in schooling fish via
  leading-edge vortex enhancement. Bioinspir Biomim 17(6):066,020

\bibitem[{Shinde and Arakeri(2013)}]{shinde2013jet}
Shinde SY, Arakeri JH (2013) Jet meandering by a foil pitching in quiescent
  fluid. Phys Fluids 25(4):041,701

\bibitem[{Shyy et~al(2010)Shyy, Aono, Chimakurthi, Trizila, Kang, Cesnik, and
  Liu}]{shyy2010recent}
Shyy W, Aono H, Chimakurthi SK, et~al (2010) Recent progress in flapping wing
  aerodynamics and aeroelasticity. Prog Aerosp Sci 46(7):284--327

\bibitem[{Triantafyllou et~al(2000)Triantafyllou, Triantafyllou, and
  Yue}]{triantafyllou2000hydrodynamics}
Triantafyllou MS, Triantafyllou GS, Yue DKP (2000) Hydrodynamics of fishlike
  swimming. Annu Rev Fluid Mech 32(1):33--53

\bibitem[{Wang(2005)}]{wang2005dissecting}
Wang ZJ (2005) Dissecting insect flight. Annu Rev Fluid Mech 37:183--210

\bibitem[{Williamson and Govardhan(2004)}]{williamson2004vortex}
Williamson CHK, Govardhan R (2004) Vortex-induced vibrations. Annu Rev Fluid
  Mech 36:413--455

\bibitem[{Xiao and Zhu(2014)}]{xiao2014review}
Xiao Q, Zhu Q (2014) A review on flow energy harvesters based on flapping
  foils. J Fluids Struct 46:174--191

\bibitem[{Young et~al(2014)Young, Lai, and Platzer}]{young2014review}
Young J, Lai JCS, Platzer MF (2014) A review of progress and challenges in
  flapping foil power generation. Prog Aerosp Sci 67:2--28

\bibitem[{Zhang et~al(2015)Zhang, Hedrick, and Mittal}]{zhang2015centripetal}
Zhang C, Hedrick TL, Mittal R (2015) Centripetal acceleration reaction: an
  effective and robust mechanism for flapping flight in insects. PloS One
  10(8):e0132,093

\bibitem[{Zhong et~al(2021)Zhong, Zhu, Fish, Kerr, Downs, Bart-Smith, and
  Quinn}]{zhong2021tunable}
Zhong Q, Zhu J, Fish FE, et~al (2021) Tunable stiffness enables fast and
  efficient swimming in fish-like robots. Sci Robot 6(57):eabe4088

\bibitem[{Zhu et~al(2019)Zhu, White, Wainwright, Di~Santo, Lauder, and
  Bart-Smith}]{zhu2019tuna}
Zhu J, White C, Wainwright DK, et~al (2019) Tuna robotics: A high-frequency
  experimental platform exploring the performance space of swimming fishes. Sci
  Robot 4(34):eaax4615

\bibitem[{Zhu et~al(2020)Zhu, Su, and Breuer}]{zhu2020nonlinear}
Zhu Y, Su Y, Breuer K (2020) Nonlinear flow-induced instability of an
  elastically mounted pitching wing. J Fluid Mech 899:A35

\bibitem[{Zhu et~al(2021)Zhu, Mathai, and Breuer}]{zhu2021nonlinear}
Zhu Y, Mathai V, Breuer K (2021) Nonlinear fluid damping of elastically mounted
  pitching wings in quiescent water. J Fluid Mech 923:R2

\end{thebibliography}

\end{document}